\documentclass[preprint,12pt,numafflabel]{elsarticle}
\usepackage[utf8]{inputenc} 
\usepackage[T1]{fontenc}
\usepackage{amssymb}
\usepackage{amsmath}
\usepackage{graphicx} % 
\RequirePackage{amsmath}
\usepackage{xcolor}
\usepackage{mathtools} 
\renewcommand{\vec}[1]{\mathbf{#1}}

\usepackage{cleveref}
\begin{document}

%\preprint{APS/123-QED}

\title{A Deterministic Ionization Algorithm for the OSIRIS Particle-in-Cell Framework 
}%
\author[1]{S. DiIorio}
\author[2,3]{R. Fonseca}
\author[4]{F. Tsung}
\author[4]{B.J. Winjum}
\author[1]{A. G. R. Thomas}
 \affiliation[1]
{organization={G\'erard Mourou Center for Ultrafast Optical Science, University of Michigan},%Department and Organization
            addressline={2200 Bonisteel Boulevard}, 
            city={Ann Arbor},
            postcode={48109}, 
            state={Michigan},
            country={USA}}
\affiliation[2]
{organization={GoLP/IPFN, Instituto Superior T\'ecnico, Universidade de Lisboa}, 1049-001 Lisbon, Portugal}
\affiliation[3]{ISCTE - Lisbon University Institute, Lisbon, Portugal}
\affiliation[4]{Department of Physics and Astronomy, University of California, Los Angeles, CA 95095, USA}
% \author{Stephen DiIorio}
%  % \altaffiliation[Also at ]{Department of Mechanical Engineering, University of Michigan, Ann Arbor, MI 48109 USA.}
% \author{Alexander G. R. Thomas}%
%  \email{Contact author: agrt@umich.edu}
% \affiliation[1]{%
%   Gérard Mourou Center for Ultrafast Optical Science, University of Michigan, Ann Arbor, MI 48109 USA\\
%   Department of Physics, University of Michigan, Ann Arbor, MI 48109 USA
% }%

\date{\today}% It is always \today, today,

\begin{abstract}
Ionization is critical in the formation and evolution of plasma dynamics; collisional ionization, in particular, is an often overlooked source of electrons when dealing with laser-plasma interactions. Ionization plays a crucial role in understanding the complex plasma kinetics, ranging from cold and sparse astrophysical settings to hot and dense fusion systems. In this paper, we describe the underlying theory for and development, validation, and verification of an extension to the standard particle-in-cell method to include a deterministic algorithm for collisional ionization physics. This algorithm offers improved accuracy, achieving up to two orders of magnitude decrease in the error of the ionization rate calculations, scales linearly in execution time with the number of macro-particles per cell, has been tested for physical correctness and benchmarked against several codes.
\end{abstract}

\maketitle
\clearpage
\setcounter{tocdepth}{2}
\tableofcontents

\section{Introduction}%
\label{sec:intro}%
Ionization is of interest for a wide range of scenarios, including astrophysical settings~\citep{sutherlandCoolingFunctionsLowDensity1993,arnouldRprocessStellarNucleosynthesis2007,marcowithMultiscaleSimulationsParticle2020}, fusion~\citep{harmsPrinciplesFusionEnergy2000,kempLaserPlasmaInteractions2014,robinsonTheoryFastElectron2014}, ion acceleration~\citep{allenDirectExperimentalEvidence2004,afshariRoleCollisionalIonization2022}, ionization-induced defocusing~\citep{raeIonizationinducedDefocusingIntense1993}, energy transport and stopping power~\citep{peterEnergyLossHeavy1991,norreysFastElectronEnergy2014}, instability modeling~\citep{ruyerGrowthConcomitantLaserdriven2020}, solid-density plasmas~\citep{wuMonteCarloApproach2017,wuParticleincellSimulationsLaser2018}, and cathode, silicon detectors, and rf plasma modeling~\citep{burgerElasticCollisionsSimulating1967,boeufMonteCarloAnalysis1982,boswellSelfConsistentSimulation1988,venderNumericalModelingLowpressure1990,venderNumericalStudiesLow1990,vahediMonteCarloCollision1995,bichselMethodImproveTracking2006}, to name but a few.

Astrophysical plasmas are generally composed of completely ionized hydrogen and helium atoms. Ionized electrons play a crucial role in the general cooling of the plasmas and also help diagnose the other elements and conditions present in nebulae and coronae. Collisional ionization gives rise to unique spectral lines through the ionization of inner shell electrons that can then be measured and compared against models to diagnose their composition~\citep{culhaneXraySpectraSolar1981,winklerSurveyXrayLine1981}.

In fusion diagnostics, elements whose spectral behavior and ionization rates are well understood are injected into the fusion plasma to serve as a diagnostic for the plasma conditions. Silicon frequently serves as this tool, and fluctuations in its measured X-ray emission can indicate sources of energy loss within the system~\citep{petrassoFullyIonizedTotal1982}. Another such element is krypton, which serves as a plasma diagnostic used in Tokamak divertors to assist with radiative cooling~\citep{lochElectronimpactIonizationAll2002}. Plasmas created in magnetic confinement fusion schemes are generally low in density but quite hot (the plasmas of their divertors, on the other hand, are usually estimated to be relatively colder). 

\begin{figure}[!htp]
  \centering
\includegraphics[width=0.75\textwidth]{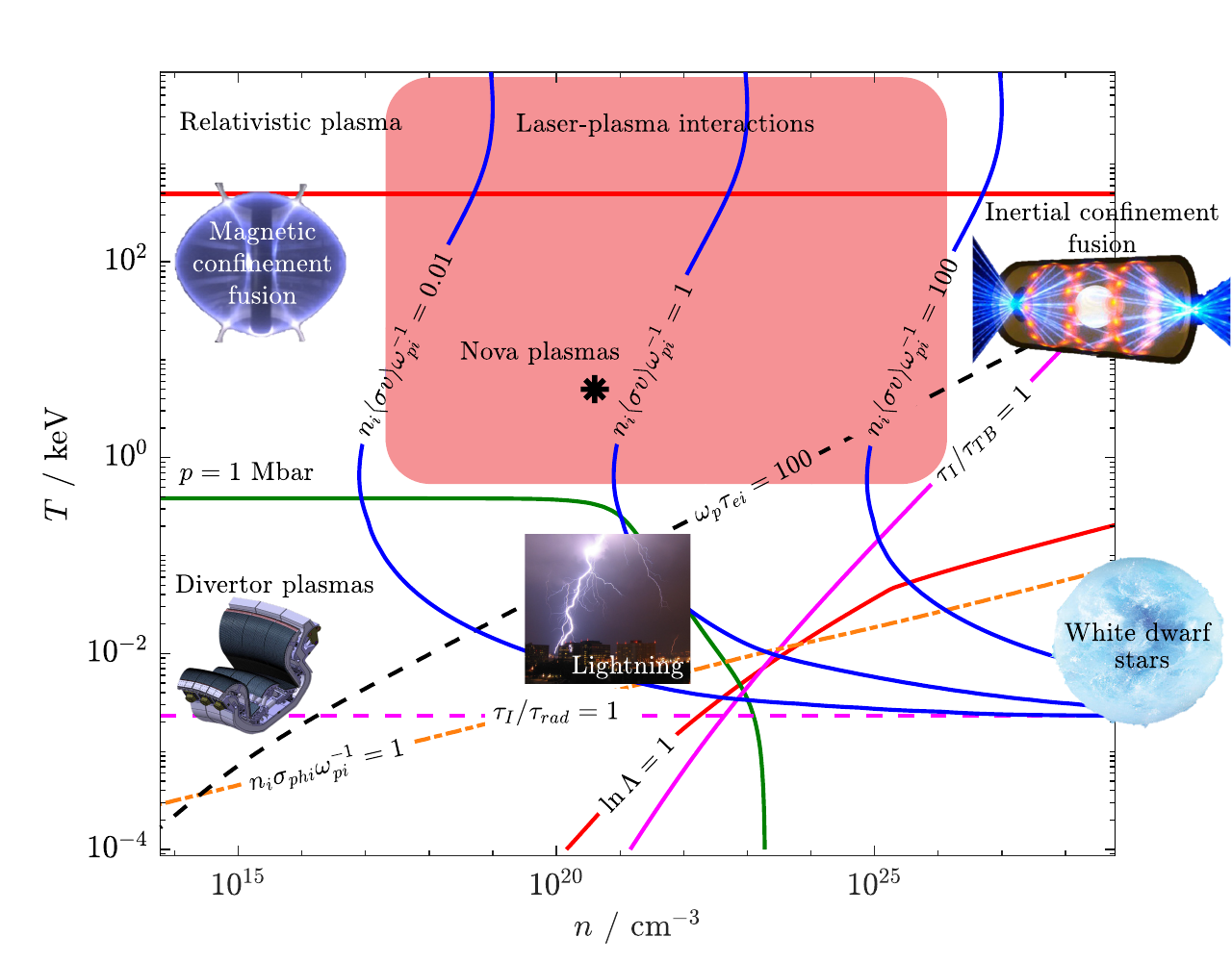}
  \caption{Plasma properties as a function of temperature and density. }
\label{chap:introduction:fig:phase_space}
\end{figure}

Consider \cref{chap:introduction:fig:phase_space}, where we show a typical phase space diagram for different plasma regimes as a function of temperature and density. In this plot, the solid magenta line \(\tau_I/\tau_{TB}\) marks where the ratio of timescales for collisional ionization and three-body recombination is equal. The dashed magenta line $\tau_I/\tau_{rad}$ marks where the timescales for collisional ionization and radiative recombination are equal. The dashed orange line shows the ratio of the photoionization timescale to the ion plasma frequency, assuming thermodynamic equilibrium. The blue lines show different contours for the ratio of the rates of collisional ionization to the ion plasma frequency. All the curves here are calculated for atomic number \(Z=1\). Laser-plasma interactions are generally hot and dense, up to plasma conditions near the hohlraum walls used in inertial confinement fusion at NIF and further increasing the density to the conditions of the compressed fusion fuel. The ionization physics is complex in these extreme cases and must be computed by considering the full detailed atomic transition kinetics. At slightly lower densities, laser-generated plasmas are often generated where collisional ionization effects start to become important, for example, with timescales comparable to the characteristic ion response timescale of the plasma (the inverse ion plasma frequency $\omega_{pi}$). Often, it is assumed that the laser intensity is sufficient to field ionize the plasma, but in reality, regions just outside the laser spot or within dense material where the laser cannot penetrate are ionized by hot electron flows. Creating appropriate numerical tools for modeling ionization physics, in particular both field and collisional ionization, is crucial to accurate kinetic plasma modeling in these regimes. 

Particle-in-cell (PIC) codes~\citep{dawsonParticleSimulationPlasmas1983,hockneyComputerSimulationUsing1988,birdsallPlasmaPhysicsComputer1991,birdsallParticleincellChargedparticleSimulations1991} are an essential tool in understanding a wide range of plasma scenarios. Very early on, the need for algorithms to be able to represent the short range collisions that is missing from the basic PIC method was identified and implemented by considering binary interactions between macroparticles (particle-particle interactions) \cite{burgerElasticCollisionsSimulating1967,shannyOneDimensionalModelLorentz1967,eastwoodP3M3DPThreedimensionalPeriodic1980,takizukaBinaryCollisionModel1977,procassiniRelativisticMonteCarlo1987} or as an effective force (particle-grid interactions) \cite{jonesGridBasedCoulombCollision1996,lemonsSmallangleCoulombCollision2009,manheimerLangevinRepresentationCoulomb1997,cohenGridbasedBinaryModel2013}. While these methodologies were primarily developed to simulate Coulomb scattering events, the same algorithms may be applied to simulating collisional ionization events. For the particle-particle type method, Ref.~\cite{perezImprovedModelingRelativistic2012} refined the Coulomb scattering models of Refs.~\cite{nanbuWeightedParticlesCoulomb1998,sentokuNumericalMethodsParticle2008} to include collisional ionization and has been implemented in several PIC codes~\cite{arberContemporaryParticleincellApproach2015,derouillatSmileiCollaborativeOpensource2018}. Inaccuracies have been noted when there is a large discrepancy in the weights and energies of the macro-particle pair colliding and how to fix them~\cite{higginsonCorrectedMethodCoulomb2020}. The PIC code \textsc{Smilei}~\citep{derouillatSmileiCollaborativeOpensource2018} uses this corrected version. Ref.~\cite{morrisImprovementsCollisionalIonization2022} further develops Higginson's corrections by improving the particle pairing process when there is a large discrepancy in the number of macro-particles of the two colliding species and this is also included in the latest versions of the PIC code \textsc{Epoch}~\cite{arberContemporaryParticleincellApproach2015}.

In this paper, a new algorithm for simulating collisional ionization within a PIC framework based on a particle-grid rather than particle-particle interaction is demonstrated. Our method treats ionization events deterministically as each particle’s contribution to the ionization rate is deposited onto a grid. This grid of ionization rates is then used to advance ion charge states. Additionally, the energy is lost per cell due to ionization physics is tracked. We interpolate this information back onto the particles; this allows for a continuous decrease in the energy of macro-particles as they participate in ionization events and allows for the easy calculation of the new momentum of ionized electrons. This algorithm is implemented within the PIC code \textsc{Osiris}~\cite{fonsecaOSIRISThreeDimensionalFully2002}. We demonstrate several tests and benchmarks against particle-in-cell codes that, when differences in the fundamental cross sections are accounted for, are in excellent agreement.

 \subsection{Theoretical background}%
%  \label{chap:theory:ssec:mrbeb}%
%  % \subsection{Modified-RBEB Cross Section}%
% % \label{chap:theory:ssec:mrbeb}%
The impact ionization cross sections used in this implementation are the Modified Relativistic Binary Encounter Bethe (MRBEB) cross sections \cite{kimExtensionBinaryencounterdipoleModel2000,guerraModifiedBinaryEncounter2012}, which match experimental electron impact ionization cross sections for a wide range of energies. For completeness, we first review the important details of the model and the sources of data. 
First, we define \(N_{k}\) and \(B_{k}\) to be the electron occupation number and electron binding energy, respectively, of subshell \(k\) and \(T\) and \(W\) to be the kinetic energy of the incident and secondary electron, respectively. We further normalize these terms as follows, defining \({w=W/B_{k}}\), \({t=T/B_{k}}\), \({b^{\prime}=B_{k}/m_{e}c^{2}}\), and \({t^{\prime}=T/m_{e}c^{2}}\). We finally define \({\beta_{b}=1-\left(1+b^{\prime}\right)^{-2}}\) and \({\beta_{t}=1-\left(1+t^{\prime}\right)^{-2}}\). We use the differential ionization cross section initially presented in Ref.~\cite{kimExtensionBinaryencounterdipoleModel2000} and modified in Ref.~\cite{guerraModifiedBinaryEncounter2012}:
\begin{equation}
  \frac{d\sigma_{k}}{dw}=\sigma_{k}^{0}\left[A_{3}f_{3}(w)+f_{2}(w)+\frac{2A_{2}}{t-1}-A_{1} f_{1}(w)\right]\;,%
  \label{chap:theory:eq:sdcs_sigma}%
\end{equation}
where
\begin{align*}
  \sigma_{k}^{0}&=\frac{4\pi a_{0}^{2}\alpha^{4}N_{k}}{\left(\beta_{t}^{2}+\chi_{n\ell j}\beta_{b}^{2}\right)2b^{\prime}}\;,\\
  f_{n}(w)&=\left(w+1\right)^{-n}+\left(t-w\right)^{-n}\;,\\
  A_{1}&=\frac{1}{t+1}\frac{1+2t^{\prime}}{\left(1+t^{\prime}/2\right)^{2}}\;,\\
  A_{2}&=\frac{t-1}{2}\frac{{b^{\prime}}^{2}}{\left(1+t^{\prime}/2\right)^{2}}\;,\;{\rm and}\\
  A_{3}&=\ln{\left(\frac{\beta_{t}^{2}}{1-\beta_{t}^{2}}\right)}-\beta_{t}^{2}-\ln{(2b^{\prime})}\;.
\end{align*}
Here, \(c\) is the speed of light in vacuum, \(m_{e}\) is the electron mass, \(\alpha\) is the fine structure constant, and \(a_{0}\) is the Bohr radius.

The \(\sigma_{k}^{0}\) term contains a coefficient that only depends on the incident electron kinetic energy and a term that depends only on \(Z\), the atomic number.%
\begin{align*}
  \chi_{n\ell j}&=2\left(\frac{C_{n\ell j}}{B_{k}}\right){\rm Ry}\;,
  % C_{n\ell j}&=0.3\frac{Z^{2}_{\text{eff}_{n\ell j}}}{2n^{2}}+0.7\frac{Z^{2}_{\text{eff}_{n^{\prime}\ell^{\prime}j^{\prime}}}}{2{n^{\prime}}^{2}}\;,\\
  % C_{n\ell j_{\text{last}}}&=\frac{Z^{2}_{\text{eff}_{n\ell j}}}{2n^{2}}\;.
\end{align*}
where \(\rm Ry\) is the Rydberg Energy, \(n\) is the principal quantum number, \(\ell \) is the orbital angular momentum, and \(j\) is the total angular momentum. The coefficients $C_{n\ell j}$ are nuclear screening constants dependent on the effective atomic number and are tabulated for elements up to \(Z=118\) in~\cite{guerraRelativisticCalculationsScreening2017}. 

\begin{figure}[t]
\centering
\includegraphics[width=\textwidth]{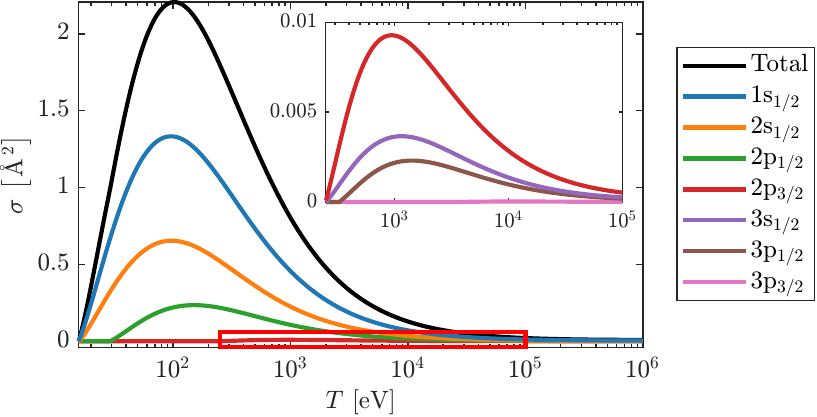}
\caption{An example of the cross sections calculated as a function of incident electron kinetic energy. This highlights the individual cross sections of each subshell of neutral argon and the total cross section obtained by summing each subshell. The region magnified by the inset is highlighted with the red box.}
\label{chap:theory:fig:argon_cs}
\end{figure}

The differential cross section is integrated and summed over all \(k\) subshell cross sections to get the total ionization cross section for an atom. Since the primary (fast) electron and the secondary (slow) electron are indistinguishable from one another, practically speaking, it is conventional to consider any electron with energy less than half the incident particle's kinetic energy (minus the binding energy required to cause ionization) to be the newly ionized electron, resulting in the following upper limit of integration.
\begin{align}
  \sigma_{k}&=\int\limits_{0}^{(t-1)/2}\frac{d\sigma_{k}}{dw}{dw}
  =\sigma_{k}^{0}\left[\frac{A_{3}}{2}\left(1-\frac{1}{t^{2}}\right)+1-\frac{1}{t}+A_{2}-A_{1}\ln{t}\right]\\
  \overline{\sigma}&=\sum_{k}\sigma_{k}%
  \label{chap:theory:eq:sigma}%
\end{align}
An example of the total cross section and the cross section of the subshells for argon is shown in \cref{chap:theory:fig:argon_cs}.

The first moment of the differential ionization cross section provides the statistical mean of \(w\), which can be used to calculate the average energy lost by the incident electron and gained by the secondary. For completeness, the expressions given in~\cite{perezImprovedModelingRelativistic2012} are
\begin{align}
 & \langle w\rangle_{k}=\frac{1}{\sigma_{k}}\int\limits_{0}^{(t-1)/2}w\frac{d\sigma_{k}}{dw}{dw}\nonumber\\
 % \begin{split}
    &=\frac{\sigma_{k}^{0}}{\sigma_{k}}\left\{
    \frac{A_{3}}{2}\frac{\left(t-1\right)^{2}}{t\left(t+1\right)}+2\ln{\left(\frac{t+1}{2\sqrt{t}}\right)}+A_{2}\frac{t-1}{4}-A_{1}\left[t\ln{t}-\left(t+1\right)\ln{\left(\frac{t+1}{2}\right)}\right]\right \}\;,%
    %     &=\frac{\sigma_{k}^{0}}{\sigma_{k}}\left\{
    % \frac{A_{3}}{2}\frac{\left(t-1\right)^{2}}{t\left(t+1\right)}+2\ln{\left(\frac{t+1}{2}\right)}-\ln{t}+A_{2}\frac{t-1}{4}
    % -A_{1}\left[t\ln{t}-\left(t+1\right)\ln{\left(\frac{t+1}{2}\right)}\right]\right \}\;,%
    \label{chap:theory:eq:e_transf_avg}
 % \end{split}
\end{align}
which is used to determine the average energy transferred to the secondary electron in units of $m_ec^2$ (note a factor of \(t\) present in \cref{chap:theory:eq:e_transf_avg} that is missing in the original reference)
\begin{equation}
  \overline{w}=\sum_{k}\langle w\rangle_{k} b^\prime_{k}\sigma_{k}/\overline{\sigma}\;,%
  \label{chap:theory:eq:e_transf}%
\end{equation}
and the average energy lost by the incident electron is
\begin{equation}
  \overline{\varepsilon}=\sum_{k}(\langle w\rangle_{k}+1)b^\prime_{k}\sigma_{k}/\overline{\sigma}\;.%
  \label{chap:theory:eq:e_lost}%
\end{equation}
It is important to note that because \(\sigma_{k}^{0}\) is a function of shell occupancy number, these values need to be calculated for each different charge state of the ion.

\begin{figure}[!htbp]
\centering
\includegraphics[width=\textwidth]{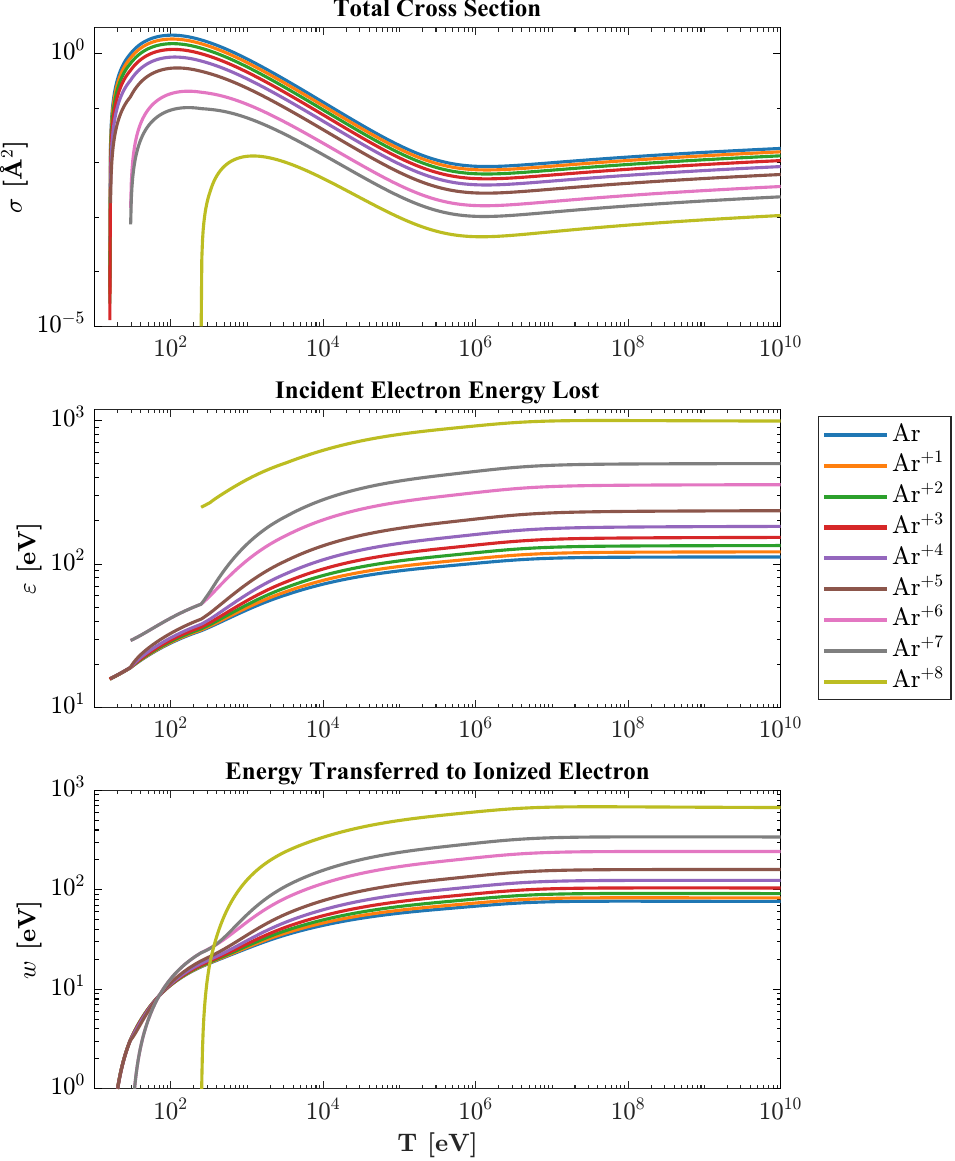}
\caption{The total cross sections for electron-impact ionization for different charge states of argon, ranging from neutral to +8. Also shown are the average energy lost by electrons and energy transferred to newly ionized electrons for those same charge states. These curves are an example calculation. % of \cref{chap:theory:eq:sigma,chap:theory:eq:e_transf,chap:theory:eq:e_lost}. We show 
Only the first eight charge states of argon are shown.% because these are the curves of interest for our discussion in \cref{chap:eprobe}.
}
\label{chap:theory:fig:argon_cs_engs}
\end{figure}

%While simpler than its predecessor, \cref{chap:theory:eq:sigma} still relies on data specific to the species at hand. In particular, 
To calculate the cross section for an atom, one must know the electron occupancy numbers of each of its subshells (and the occupancy of each of its ionization states, which can be nontrivial), the binding energies of each of its electrons, and the effective nuclear charge--effective because there are shielding effects from the other electrons present--each electron experiences. Both experimental and computational work has gone into creating reference tables for different atomic binding energies~\cite{beardenReevaluationXRayAtomic1967,carlsonCalculatedIonizationPotentials1970,lotzElectronBindingEnergies1970,luRelativisticHartreeFockSlaterEigenvalues1971,desclauxRelativisticDiracFockExpectation1973,carlsonPhotoelectronAugerSpectroscopy1975,larkinsSemiempiricalAugerelectronEnergies1977,shirleyCoreelectronBindingEnergies1977,sevierAtomicElectronBinding1979,bruchKshellBindingEnergies1985,rodriguesSystematicCalculationTotal2004,thompsonXrayDataBooklet2009,kramidaNISTAtomicSpectra2022}, screening constants~\cite{slaterAtomicShieldingConstants1930,clementiAtomicScreeningConstants1963,clementiAtomicScreeningConstants1967,guerraRelativisticCalculationsScreening2017}, and shell occupation number~\cite{luRelativisticHartreeFockSlaterEigenvalues1971,desclauxRelativisticDiracFockExpectation1973,rodriguesSystematicCalculationTotal2004,kramidaNISTAtomicSpectra2022}. With the variety of databases available, work has gone into determining which of the binding energy values produce the most accurate replication of experimental cross-section data~\cite{piaEvaluationAtomicElectron2011}. Here, the binding energies from \cite{carlsonPhotoelectronAugerSpectroscopy1975} are used, the most recent calculations of screening constants from \cite{guerraRelativisticCalculationsScreening2017}, and electron shell occupation number from the NIST Atomic Spectra Database~\cite{kramidaNISTAtomicSpectra2022}. These choices used with \cref{chap:theory:eq:sigma} have led to good agreement with other experimental and theoretical cross-section results~\cite{powellCrossSectionsIonization1976,bellRecommendedDataElectron1983,deutschCalculationAbsoluteElectron1987,tawaraTotalPartialIonization1987,freundCrosssectionMeasurementsElectronimpact1990}.

In the computational model, to model the momentum balance, we consider the interaction of an average electron. After an ionizing collision, the new momentum $\vec{p}_e^\prime$ for the average incident electron, assuming that the momentum transfer is collinear with the original momentum $\vec{p}_e$, is 
\begin{equation}
\label{chap:implementation:eq:comp_p_lost}%
\vec{p}_{e}^{\prime}=\alpha_{\varepsilon}\vec{p}_{e}+\left[(1-\alpha_{\varepsilon})\gamma_{e}^{*}-\overline{\varepsilon}\right]\frac{m_{e}}{m_{i}}\vec{p}_{i}\;,
  \end{equation}
 which makes use of the normalized electron energy expressed in the ion's rest frame, \(\gamma_{e}^{*}=\gamma_{e}\gamma_{i}-\vec{p}_{e}
\cdot\vec{p}_{i}/(m_{e}m_{i}c^2)\) and the ratio of the momenta of the electron before and after the interaction in the ion rest frame
\begin{equation}
\alpha_{\varepsilon}=\frac{|\vec{p}_{e}^\prime|}{|\vec{p}_{e}|}=\frac{\sqrt{(\gamma_{e}^*-\overline{\varepsilon})^{2}-1}}{\sqrt{\gamma_e^{*2}-1}}\;.\label{eqn:alpha_e_original}
\end{equation}
 Likewise, the momentum of the average secondary electron generated by ionization, again assuming that it is collinear with the initial ionizing electron, is
  \begin{equation}
\label{chap:implementation:eq:comp_p_transf}
\vec{p}_{w}=\alpha_{w}\vec{p}_{e}+\left(\overline{w}+1-\alpha_{w}\gamma_{e}^{*}\right)\frac{m_{e}}{m_{i}}\vec{p}_{i}\;,
\end{equation}
where 
\begin{equation}
  \alpha_{w}=\frac{|\vec{p}_{w}|}{|\vec{p}_{e}|}=\frac{\sqrt{\overline{w}(2+\overline{w})}}{\sqrt{\gamma_{e}^{*2}-1}}\;.\label{eqn:alphaw}
\end{equation}
Note that $\alpha_\varepsilon\in[0,1],\;\alpha_w\in[0,1)$.
For non-relativistic ions with $\vec{p}_i\lesssim\vec{p}_e$, these expressions reduce to $\gamma_e^*\simeq\gamma_e$, $\vec{p}^\prime_e=\alpha_\varepsilon \vec{p}_e+\mathcal{O}(m_e/m_i)$ and $\vec{p}^\prime_w=\alpha_\varepsilon \vec{p}_w+\mathcal{O}(m_e/m_i)$. In the implementation discussed here, we take this non-relativistic ion limit for simplicity, which also means that the cross-section rate calculations, which are described in the above expressions also in the ion rest frame, maybe used as is without transforming between frames.
% Additionally, our previous definitions of \(\alpha_{\varepsilon}\) and \(\alpha_{w}\) used in %\cref{chap:implementation:eq:momlost,chap:implementation:eq:momtransf_mobile}
% \cref{eqn:alpha_e,eqn:alphaw}
% and also used here are slightly modified to use \(\gamma_{e}^{*}\) instead of just \(\gamma_{e}\). 
\subsection{Field Ionization Rates}%
\label{chap:implementation:sssec:fi_rates}%
%
%Implementing some field ionization routine within the basic PIC infrastructure is often simple. The electric fields calculated during every PIC loop determine the local ionization rates throughout the simulation. In other PIC codes, the local electric field is interpolated onto each ionizable macro-particle and used to calculate some probability of that particle ionizing. 
% This is done utilizing ionization rates proposed by \cite{ammosovTunnelIonizationComplex1986}, commonly referred to as the ADK model.
% {\color{red}
% Unlike \textsc{Smilei} or \textsc{Epoch}, however, \textsc{Osiris} calculates these ionization rates as a grid quantity. Additionally, they instead use the formulation provided by \cite{bruhwilerParticleincellSimulationsTunneling2003}, which offers a simplification to the rates proposed in the ADK model. 
Field ionization is implemented assuming that the ionizing electric field, whether from laser light or arising from the generated plasma fields, is in the tunneling regime (small Keldysh parameter \cite{keldyshIonizationFieldStrong1965}). For calculating rates of ionization, variants of the ADK model \cite{ammosovTunnelIonizationComplex1986} are typically used. Since the laser cycle is resolved in the PIC code, in this work an instantaneous rate (non cycle-averaged, as in ADK) is used, 
in the form of an equation for the tunneling ionization rate of the outermost electron of an atom in ionization state \(i\) as a function of the local electric field strength \cite{bruhwilerParticleincellSimulationsTunneling2003}, % ~\cite[Eqn.~3 in Ref.][]{bruhwilerParticleincellSimulationsTunneling2003}:
\begin{equation}
  \mathsf{R}_{\text{TI}}\left[{\rm s^{-1}}\right]\approx{1.52\times10^{15}}\frac{4^{n^{*}}\xi_{i}}{n^{*}\Gamma(2n^{*})}\left(20.5\frac{\xi_{i}^{3/2}}{E}\right)^{2n^{*}-1}\exp\left(-6.83\frac{\xi_{i}^{3/2}}{E}\right)\;.%
  \label{chap:implementation:eq:bruh_rate}%
\end{equation}
\(E\) is the local electric field strength here in units of GV/m, \(\xi_{i}\) is the energy of electron's unperturbed ground state in units of eV, \(\Gamma\) is the Gamma function, and we approximate the effective principal quantum number as \(n^{*}\approx3.69Z/\xi_{i}^{1/2}\), where \(Z\) is the charge state of the ion after ionization. %Before this body of work, this was the extent of field ionization calculated within \textsc{Osiris}.

%We cast \cref{chap:implementation:eq:bruh_rate} into the more convenient form of \(\mathsf{R}_{\text{TI}}=A\times E^{-C}\exp(-B/E)\) where \(A\), \(B\), and \(C\) are just numerical coefficients dependent on the binding energy and effective principal quantum number. %In the usual operating procedure, a user of the code was expected to calculate the \(A\), \(B\), and \(C\) coefficients for each ionization state they were interested in. \textsc{Osiris} provided precalculated rates for a handful of elements users could select. Since then, we streamlined this process so that 
%Users  need only to specify the atomic number of the element and the  coefficients and ionization states are automatically calculated using the ionization energies tabulated from~\cite{kramidaNISTAtomicSpectra2022}. Users can still provide their own coefficients if they wish to use their own rates.

Tunneling rates, such as Ref.~ \cite{bruhwilerParticleincellSimulationsTunneling2003} are only valid for field magnitudes below the critical field strength
\begin{equation}
E_{\text{cr}}=E_{a}\left(\sqrt{2}-1\right)\left|\frac{\xi_{i}}{2I_{H}}\right|^{\frac{3}{2}}\;,
\label{chap:implementation:eq:critfield}%
\end{equation}
where \({I_{H}=m_{e}e^{4}/\left(2\hbar^{2}\right)\approx{13.6}}\) eV is the ionization potential of hydrogen, and the characteristic atomic field is \({E_{a}=m_{e}^{2}e^{5}\hbar^{-4}\approx{510}}\) GV/m. This field strength marks the boundary between tunneling ionization and barrier-suppression ionization, where an electron can classically escape the atomic potential. %In fact, within \textsc{Osiris}, nothing particularly sophisticated was done when calculating the ionization rates; \cref{chap:implementation:eq:bruh_rate} was used exclusively regardless of how strong or weak the electric field actually was. This has the potential to greatly underestimate the amount of ionization that can occur once we start dealing with field strengths greater than the critical strength.

Several studies have extended the theory of field ionization to transition to the barrier suppression regime~\cite{krainovIonizationRatesEnergy1997,posthumusMolecularDissociativeIonisation1997}. In particular, Ref.~\cite{kostyukovFieldIonizationShort2018} derived field ionization rates using both classical and quantum analytic considerations. Their model encompasses a range of field values, including extremely strong electric fields where barrier-suppression ionization, a poorly explored regime, is dominant, and they derive the following rate equation:
\begin{equation}
  \mathsf{R}_{\text{BSI}}(E)\approx0.8\omega_{a}\frac{E}{E_{a}}\sqrt{\frac{I_{H}}{\xi_{i}}}\;.%
  \label{chap:implementation:eq:bsi_rate}%
\end{equation}
\({\omega_{a}=m_{e}e^{4}\hbar^{-3}\approx{4.1\times10^{16}}\;{\rm s^{-1}}}\) is the inverse atomic time interval.
%As the authors\citeauthor{kostyukovFieldIonizationShort2018} state, the large number of formulas at our disposal, which cover a variety of different ranges, means that we should try to introduce a single equation which is ``simple and computationally cheap, valid in a wide range of laser intensities, and applicable to all types of atoms as well as all ion charges.'' Their suggestion is 
% As a simple model implementationm a piecewise function is proposed such that after combining the tunneling ionization rate, \cref{chap:implementation:eq:bruh_rate}, with the BSI rate, \cref{chap:implementation:eq:bsi_rate}, and using the critical electric field as the transition point between the two regimes, we provide a more durable ionization model that encompasses a wider range of anticipated field strengths,
% \begin{equation}
%   \mathsf{R}_{\text{FI}}(E)\approx\
%   \begin{cases}
%   \mathsf{R}_{\text{TI}}(E), &|E|\leq E_{\text{cr}},\\
%   \mathsf{R}_{\text{BSI}}(E), &|E|>E_{\text{cr}}
%   \end{cases}\;.%
%   \label{chap:implementation:eq:complete_firate}%
% \end{equation}
% \subsubsection*{On an \emph{even more} thorough field ionization model}%
% \label{chap:implementation:sssec:thorough_fi_model}%
%
Kostyukov and Golovanov~\cite{kostyukovFormulaIonisationRate2020}, extended this model to include an intermediary ionization rate at the transitional region where \(|E|\sim E_{\text{cr}}\), which they claim more accurately captures ionization in regions with laser fields of this strength. A simple piecewise rate equation that spans these regimes is
\begin{equation}
  \mathsf{R}_{\text{KG}}(E)\approx
  \begin{cases}
    \mathsf{R}_{\text{TI}}(E), &|E|\leq E_{1}\;,\\
    \mathsf{R}_{\text{BM}}(E), &E_{1}<|E|\leq E_{2}\;,\\
    \mathsf{R}_{\text{BSI}}(E), &|E|>E_{2}\;,
  \end{cases}%
  \label{chap:implementation:eq:ternary_fi}%
\end{equation}
where \(\mathsf{R}_{\text{BM}}\) is an ionization rate for intermediate regimes derived from an expression provided in \cite{bauerExactFieldIonization1999}
\begin{equation}
  \mathsf{R}_{\text{BM}}(E)\approx2.4\omega_{a}\left(\frac{E}{E_{a}}\right)^{2}\left(\frac{I_{H}}{\xi_{i}}\right)^{2}\;.%
  \label{chap:implementation:eq:bm_rate}%
\end{equation}

This piecewise approach to field ionization rates requires that the ionization rate is a continuous function. In other words, \(E_{1}\) and \(E_{2}\), the field transition thresholds, must satisfy \(\mathsf{R}_{\text{TI}}(E_{1})=\mathsf{R}_{\text{BM}}(E_{1})\) and \(\mathsf{R}_{\text{BM}}(E_{2})=\mathsf{R}_{\text{BSI}}(E_{2})\). Considering that these rates are different for each ionization state of an atom, this determination must be made for each level. 
\(E_{2}\) can be easily calculated via \(\frac{1}{3}E_{a}\left(\frac{I_{H}}{\xi_{i}}\right)^{-3/2}\); \(E_{1}\), however, is not as straight-forward. The electric field where \(\mathsf{R}_{\text{TI}}\) and \(\mathsf{R}_{\text{BM}}\) are equivalent can be found analytically but involves the calculation of a Lambert \(W\) function. %This function, computationally at least, is not trivial to evaluate, and many programming languages do not provide internal functions to evaluate them. Only through external libraries can one evaluate these functions through iterative means or approximations. As a result, it is easier to just incrementally scan through field values and evaluate \cref{chap:implementation:eq:bruh_rate,chap:implementation:eq:bm_rate} for each ionization level until we reach their crossing point. We do this at simulation initialization and store the value for use throughout the rest of the simulation.

\begin{figure}[t]
\centering
\includegraphics[width=\textwidth]{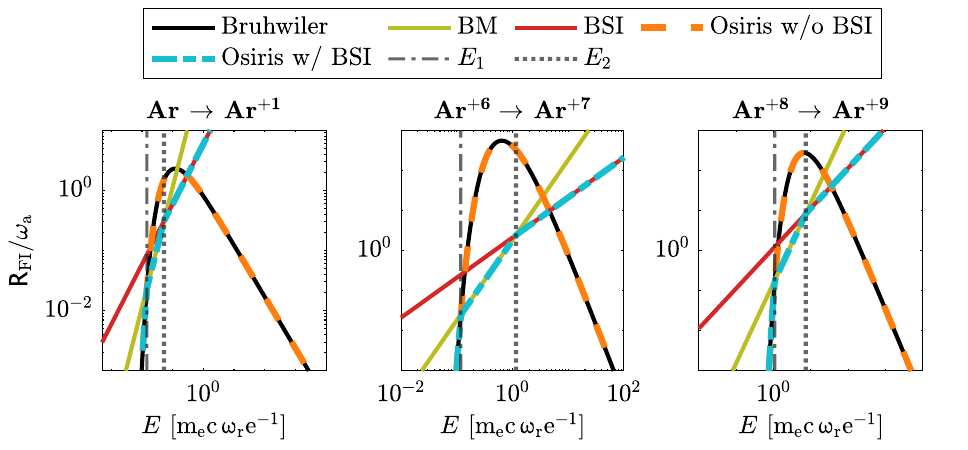}
\caption{We calculate the different field ionization rates as a function of electric field strength for various ionization states of argon. \Cref{chap:implementation:eq:bruh_rate,chap:implementation:eq:bm_rate,chap:implementation:eq:bsi_rate} were used to calculate the lines labeled Bruhwiler, BM, and BSI, respectively. Overlayed on these individual rate curves is the ionization rate used within \textsc{Osiris} when a user uses field ionization with and without BSI calculations. The vertical lines indicate the transitionary electric fields, \(E_{1}\) and \(E_{2}\), when using the BSI ionization rate.}
\label{chap:implementation:fig:osiris_fi_rate_compare}
\end{figure}

%The ionization rates we implemented were not precisely those proposed in \cite{kostyukovFormulaIonisationRate2020}. This is because the full ADK~\cite{ammosovTunnelIonizationComplex1986} rates were used, whereas we used the formulation in~\cite{bruhwilerParticleincellSimulationsTunneling2003}. This is because the ADK model is laser cycle averaged, whereas in the kinetic simulations, we require the instantaneous tunneling rate.

We show the ionization rates from these different models %from \textsc{Osiris} using either implementation 
in \cref{chap:implementation:fig:osiris_fi_rate_compare}. There are two regions of interest. For large electric field strengths compared to the critical field, the tunneling rate would dramatically underestimate the ionization rate. For intermediate field strengths, the tunneling rate would overestimate the ionization rate. %If a user solely relied on the \cref{chap:implementation:eq:bruh_rate}, we can see that the simulated ionization rates would be grossly underestimated. Alternatively, but perhaps not so consequential, is for intermediate field strengths, where the amount of ionization might be overestimated if using the tunneling rates. %in Ref.~\cite{bruhwilerParticleincellSimulationsTunneling2003} as opposed to those in Ref.~ \cite{kostyukovFormulaIonisationRate2020}. 
%This may not be so much an issue compared to the former scenario, depending on what one uses to drive the ionization dynamics. 
In practice, for optical lasers (e.g. 800 nm Ti:Sapph) the ionization is generally rapid compared to the laser period such that full ionization occurs before the critical field strength is reached and the barrier suppression considerations are relatively unimportant. With shorter wavelength light sources, such as XUV/ soft X-rays or, e.g., attosecond pulsed light, we see a large discrepancy between the different ionization models since the rate is relatively slow compared to the optical period. In general, for high-intensity laser-plasma interactions, since the plasma fields for solid densities and above may have short effective wavelengths/oscillation periods and may have strong field strengths, it is important to include these corrections.

%%%%%%%%%%%%%%%%%%%%%%%IMPLEMENTATION%%%%%%%%%%%%%%%%%%%%%%%
\section{Implementation}%
\label{sec:implementation}%
%
% Most existing PIC codes include some form of an ionization routine, and many of these open-source codes simulate ionization using the following procedure:
% \begin{enumerate}
%   \item An ionization rate, \(\mathsf{R}\), is calculated from the simulation's physical properties. These can be the electric fields in the simulations for field ionization or particle properties, as with collisional ionization.
%   \item A uniformly distributed random number between 0 and 1, \(r\), is generated for each macro-particle with a bound electron and can undergo further ionization.
%   \item A probability for each particle to be ionized is calculated, typically via a function of the form \(P=1-\exp(-\mathsf{R}\Delta t)\), and if \(r<P\) is satisfied that particular macro-particle undergoes an ionization process.
% \end{enumerate}
% Immediately obvious is the reliance of these algorithms on a Monte-Carlo process to determine if an ionization event occurs. In this chapter, we discuss the differences in approach our PIC code of choice, \textsc{Osiris}, uses to model ionization and, consequently, the algorithmic decisions necessary to simulate these physical events.
In this section we describe the implementation of an ionization algorithm in the \textsc{Osiris} particle-in-cell code based on the solutions to ionization rate equations on grids with interpolation to/from the particles.
We break the discussion of our algorithm into two basic models: the ionization of neutral objects with \emph{immobile} ions and the ionization of neutral objects with \emph{ionizable, mobile} ions. The immobile ions have a ``neutral'' object that contains data structures keeping track of the ionization of a fixed medium within the simulation. However, this model does not produce any ion macro-particles that could potentially influence the state of the system; it only produces ionized electron macro-particles. This is acceptable in some situations, e.g., simulations on short enough timescales where ion motion does not play a role or simulations with a moving window where ions leave the simulation box relatively quickly. The ionizable, mobile ions model addresses situations where users wish to produce ions in various ionization states and track their motion. 

% For completeness, there is an object already existent within \textsc{Osiris} that could be considered in-between these two options: a neutral object with mobile ions (note the lack of the keyword: \emph{ionizable}, mobile ions). This object has all the same properties as the neutrals with immobile ions except that ions can manifest as additional macro-particles within the simulation. These ion macro-particles, however, do not participate in any further ionization process and thus cannot change their ionization state. But it operates almost equivalently to the ionizable species with immobile ions, and our discussion below is just as applicable, so we do not feel it necessary to delve into any specifics further.

\subsection{Ionization Infrastructure in \textsc{Osiris}}%
\label{chap:implementation:sec:osirisinfra}%
Before going into the specifics of the differences between ionization algorithms, it is first worth a detailed explanation of the internal data structure within \textsc{Osiris} that constrained and influenced the algorithmic design choices. 

\subsubsection{Immobile Ions}%
\label{chap:implementation:ssec:immobileinfra}%

Ionizable species (referred to as `neutrals' in the \textsc{Osiris} framework) are objects that do not contain any macro-particles but contain information about the local number density of every ionization state of that species in every grid cell. A `neutral' species is nothing more than two sets of grids: one collection of computational grids, \(\overline{Z}\), describes the average ionization state profile throughout the simulation space for each ionization state of that element, and the second collection, \(\overline{\mathsf{R}}\), is the ionization rate for each ionization state throughout the simulation domain. Note that grid-based quantities will be distinguished by a bar over the variable name throughout this paper. In the input deck, the user specifies the maximum ionization state, \(z_{\text{max}}\), a neutral can reach, e.g., we could specify for helium to be ionized to either a +1 state and then no further or a maximum of +2, or we could select for aluminum to be ionizable to a maximum of any charge state between +1 and +13.

\(\overline{Z}\) and \(\overline{\mathsf{R}}\) both contain a set of \(z_{\text{max}}+1\) grids (+1 to account for a charge state of 0), and the values of all the cells in the \(\overline{\mathsf{R}}\) grids are initially zero. However, we impose a few constraints on the \(\overline{Z}\) grids. Each cell, \(j\), of each \(\overline{Z}\) grid contains a value \(\overline{Z}_{i,j}\in[0,1]\), where \(i\) is the charge state. In essence, the value of $\overline{Z}$ in a cell captures the fraction of the number density of the neutral object in that particular ionization state. During the setup of a simulation, every cell in the \(\overline{Z}_{0}\) grid (i.e., neutral atoms) is initialized with a value of 1, and the value of \(\overline{Z}_{i}\) in every other cell in the remaining grids, \(i>0\), are assigned the value 0. As the program proceeds, these values increase and decrease as we ionize the local populations.

There is an obvious constraint with an infrastructure designed this way: the sum over \(\overline{Z}_{i}\) in the \(j^{\text{th}}\) cell must be 1. In other words,
\begin{equation*}
  \sum_{i=0}^{z_{\text{max}}}\overline{Z}_{i,j}=1\;.
\end{equation*}
The user defined density profile itself is introduced through the gridded quantity \(\overline{Z}_{\textsc{Prof}}\). %The user specifies a profile that is discretized onto this grid. 
Cell values, in this case, are \(\overline{Z}_{\textsc{Prof}}\geq 0\) and static for the duration of the simulation. Subsequently, to get the number density for any given charge state of the ion at any timestep,
\begin{equation}
  n_{i}=\overline{Z}_{i} \overline{Z}_{\textsc{Prof}}\;.%
\label{chap:implementation:eq:density_profile}%
\end{equation}

A grid is created to keep track of the degree of ionization, with its cells being limited to values between 0 and \(z_{\text{max}}\). The total amount of charge in each cell is calculated every timestep as follows:
\begin{equation}
  \overline{Z}^{*}=\sum_{i=1}^{z_{\text{max}}}i\overline{Z}_{i}\;.%
  \label{chap:implementation:eq:total_charge}%
\end{equation}
We use this grid when determining whether to inject new, ionized macro-particles into the simulation (discussed in \cref{chap:implementation:sssec:immobile_inject}) and as a diagnostic. Solely in the case of collisional ionization do we create another set of grids that collects information regarding the momentum to transfer to newly ionized macro-particles. How we use this grid is explored more fully in \cref{chap:implementation:sssec:ci_rate_immobile}.

Finally, associated with a neutral is a separate species that stores information about the secondary particles generated during an ionization event, i.e., electrons. This species contains macro-particles that are free to move and influence the state of the system at later timesteps.

\subsubsection{Mobile Ions}%
\label{chap:implementation:ssec:mobileinfra}%
The new infrastructure developed in this paper, allowing ionizable moving ions, mostly does away with the grid-based approach and shifts much of the information to the macro-particles themselves. At simulation initialization, we create \(z_{\text{max}}+1\) species that are associated with different charge states of the neutral object, ranging from a charge state of zero to the maximum desired ionization level. Each species has with it a set of three new variables. The first, \(\zeta\), measures how ionized each macro-particle is between a species' current ionization level and its next. If \(z^{*}\), a continuous value between 0 and \(z_{\text{max}}\), is the ionization state of the particle, we define
\begin{equation}
  \zeta=1-\left(z^{*}-\left\lfloor z^{*}\right\rfloor\right)\;.%
  \label{chap:implementation:eq:zeta}%
\end{equation}

The second variable is a random number between \((0,1)\) (non-inclusive), which acts as a promotion threshold for that particular particle. Once a particle's \(\zeta\) value reaches that threshold, we promote that particle to the species corresponding to the next ionization state. The third variable stores information regarding the momentum to be transferred to a newly ionized electron once the ion macro-particle is promoted, or ionized, from one state to the next. During setup, we assign only the species for charge state zero (the neutral) macro-particles to correctly build the density profile desired. These particles are all assigned a value of \(\zeta=1\), a random promotion threshold, and initial values of zero for momentum to be transferred. The rest of the charge states have no macro-particles at this point. 
 \(\overline{Z}_{i}\) and momentum grids are still allocated but are only used to calculate local rates and momenta that we interpolate onto the ion macro-particles themselves.

\subsection{Ionization Procedure}%
\label{chap:implementation:sec:procedure}%

\begin{figure}[t]
\centering
\includegraphics[width=\textwidth]{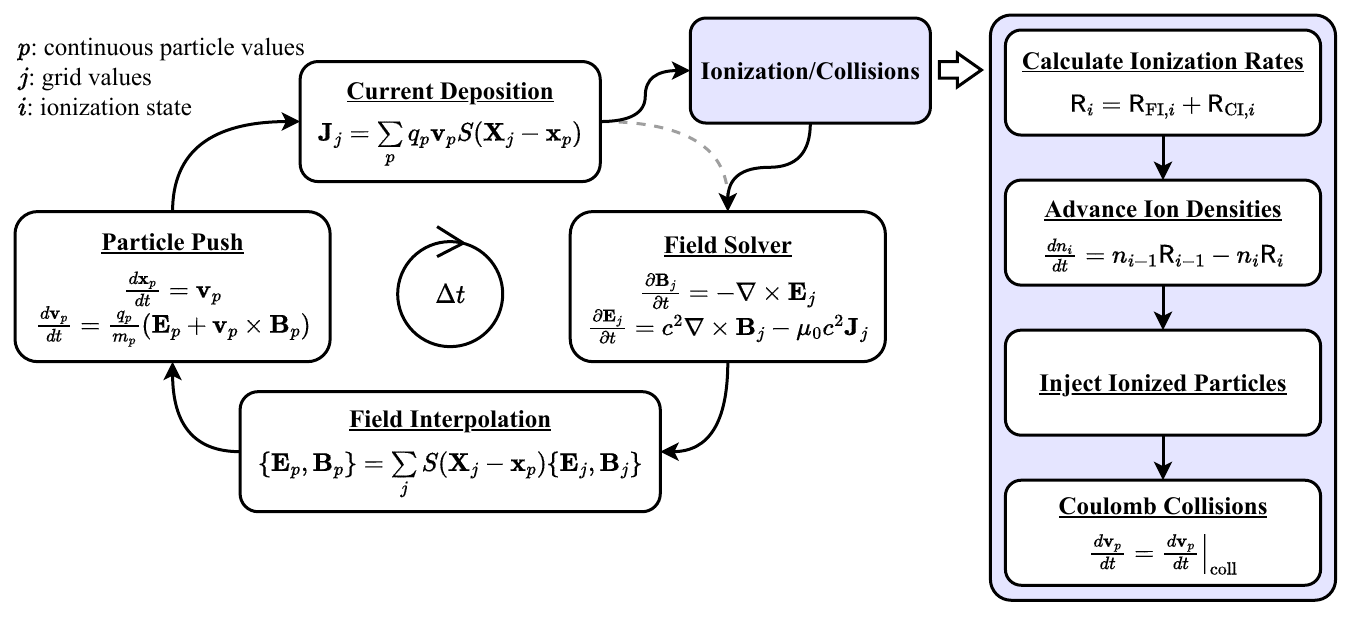}
\caption{Flow chart of the ionization and collision routines in the PIC loop. During ionization, the ionization rates are calculated and the rates used to advance the ion densities, inject new macro-particles, and, if desired, perform Coulomb collisions. This is done every timestep before returning to the usual set of operations.}
\label{chap:implementation:fig:modpicloop}
\end{figure}

In general, the ionization algorithm can be divided into three distinct processes that take place over a simulation timestep (see \cref{chap:implementation:fig:modpicloop} for a flow diagram for the modified PIC loop):
\begin{enumerate}
  \item calculation of the ionization rates and contributions to momentum lost/transferred for each specified macro-particle species on a spatial grid,
  \item deterministic advancement of the local ion densities using the calculated rates,
  \item injection of new macro-particles if enough charge has been generated with the proper handling of any momentum transfer.
\end{enumerate}

\subsubsection{Field ionization}
Field ionization is implemented in \textsc{Osiris} by first putting the tunneling ionization rate \cref{chap:implementation:eq:bruh_rate} into the more convenient form \(\mathsf{R}_{\text{TI}}=A\times E^{-C}\exp(-B/E)\) where \(A\), \(B\), and \(C\) are just numerical coefficients dependent on the binding energy and effective principal quantum number. %In the usual operating procedure, a user of the code was expected to calculate the \(A\), \(B\), and \(C\) coefficients for each ionization state they were interested in. 
The new version of \textsc{Osiris} provides pre-calculated data for all elements through the user specifying the atomic number of the element. On specifying $Z$, the coefficients for the tunneling rate and the ionization states used, e.g., in the barrier suppression rates, are then automatically calculated using the ionization energies tabulated in~\cite{kramidaNISTAtomicSpectra2022}. Users can still provide their own coefficients if they wish to use their own rates.

To find the thresholds between the functions in the piecewise model of \cref{chap:implementation:eq:ternary_fi}, as stated before the lower threshold can be calculated algebraically but for the upper threshold, many programming languages do not provide internal functions to evaluate the required special function. However, the threshold can be pre-calculated by an incremental scan through field values evaluating \cref{chap:implementation:eq:bruh_rate,chap:implementation:eq:bm_rate} for each ionization level until the crossing point is reached. This is performed at simulation initialization and the value stored for comparative use throughout the simulation.
% \subsubsection{Ionization Rates}%
% \label{chap:implementation:ssec:ion_rate}%
% {\color{red}%
% \textsc{Osiris}, and many other PIC codes, rely on calculating a rate, which is used to determine how much ionization should take place throughout the simulation. Some PIC codes (we will focus primarily on how these routines are completed in \textsc{Smilei}~\cite{derouillatSmileiCollaborativeOpensource2018} and \textsc{Epoch}~\cite{arberContemporaryParticleincellApproach2015}) calculate these rates for each macro-particle in the simulation. In the case of field ionization, the local electric field values are interpolated from their grid onto each macro-particle, and, in the case of collisional ionization, local macro-particles are paired together and share information to perform a calculation to determine the probability of ionization. This design, particularly concerning the latter scenario, has enormous computational costs. This process, however, is wholly unnecessary given the infrastructure in \textsc{Osiris}.

%In the case of neutrals, because ion densities are stored as grid quantities of their own, we desire to calculate ionization rates as a grid quantity that we can easily use to determine how much a cell should be ionized. And even though ionization information is stored as a particle value in the case of mobile ions, we still gain inspiration from their immobile counterparts to implement a more deterministic ionization method. The next question we address is how to calculate these rates.

\subsubsection*{Immobile Ions}%
\label{chap:implementation:sssec:fi_immobile}%
With this formulation in hand, we use the local electric field at every timestep to calculate the ionization rates that the neutral object undergoes. Depending on the simulation dimension and order of shape function used, some reinterpolation occurs to align the local electric field values from their Yee staggered positions to different positions with respect to a grid cell. For linear and cubic (odd order) shape functions, the magnitude of the electric field is averaged at the center of each cell \emph{volume}. For quadratic and quartic (even order) shape functions, the magnitude of the electric field is calculated at the cell corners. From here, these averaged field components are used to calculate the magnitude of the electric field that is then used in the above field ionization rate calculations. A diagram with the positions of these quantities in all three dimensions is shown in \cref{chap:implementation:fig:pic_cell_w_markers}.

\subsubsection*{Mobile Ions}%
\label{chap:implementation:sssec:fi_mobile}%
On the other hand, ionizable mobile ions take advantage of the fact that we have moved these calculations from a rate grid to the particle level. The local electric fields are now interpolated from the simulation's field grid to each ion macro-particle, as would be done when finding the local fields to use during a particle push. Then, the magnitude of that interpolated electric field is calculated and used to determine the field ionization rate for that ion macro-particle.

\subsubsection{Collisional Ionization Rates}%
\label{chap:implementation:ssec:ci_rates}%
A new addition to \textsc{Osiris} with this body of work is the consideration of collisional ionization as a source of new particles. Several PIC codes have developed collisional ionization packages prior, for example~\cite{perezImprovedModelingRelativistic2012,arberContemporaryParticleincellApproach2015,derouillatSmileiCollaborativeOpensource2018}. How to include collisional effects, though, is not as obvious as field ionization, as we will see, considering there is no equivalent to the electric field naturally calculated during the base PIC loop that we can draw information from. These codes use a Monte-Carlo approach to determine whether ionization occurs. Local particles are paired, and a random number is generated for every potential collision event. Whether this number is above or below a calculated probability determines whether or not the event occurs. While these codes can simulate collisional ionization, the fact that events are based on a random process introduces unwanted noise into the simulation. Trying to sample a thermal distribution of particles properly takes up to several thousand to several tens of thousands of particles per cell~\cite{arberContemporaryParticleincellApproach2015} to achieve results that start to overcome the noise generated by this random process. Here, we seek a deterministic algorithm by solving the rate equations, creating a smoother ionization process for low particle statistics.

Our algorithm explicitly calculates the rate at which collisional ionization should occur throughout the simulation and properly advances the relevant ion densities while injecting new macro-particles when necessary. More formally, our goal is to determine how the ion number densities, \(n_{i}\), change over time and to use the amount of charge generated through ionization to inject new macro-particles into the simulation. Doing this requires we find some means of evaluating a set of coupled differential equations that take on the form
\begin{equation}
  \frac{dn_{i}}{dt}=n_{i-1}\mathsf{R}_{i-1}-n_{i}\mathsf{R}_{i}\;,%
  \label{chap:implementation:eq:dni_dt}%
\end{equation}
where \(\mathsf{R}_{i}\) is the ionization rate per ion for ions with charge state \(i\). In general, a given ion state of a species can either grow in number density as ions in a less ionized state become more ionized or diminish in number density as it becomes more ionized. The outlying cases to this generalization are the neutral atom, which can only decrease in number density (one of our assumptions described below), and the fully ionized ion, which can only increase in number density.

We make a few assumptions within our algorithm. First, we calculate the cross sections for each ion assuming that the ion is in its ground state. Another way of stating this is that we do not consider the details of the excitation or deexcitation of atoms. The cross sections we calculate measure how likely it is for an incident electron to collide with and ionize any of the bound electrons. An ionization process could remove an electron from the K-shell, a valence electron, or any in-between. These missing electrons could influence future cross-section calculations as gaps would directly affect the occupation number, \(N_{k}\), used in \cref{chap:theory:eq:sdcs_sigma}. However, the \textsc{Osiris} infrastructure does not readily record the information necessary to keep track of all the possible excitation states without significant modifications to the core of the code. Thus, even if an inner shell electron were to be ionized via this process (creating a hollow atom), we assume that the ion relaxes rapidly to its ground state.

Second, we ignore any recombination effects. This includes radiative recombination, dielectric recombination, and three-body recombination. In general, recombination effects are unlikely to play an important role for a few picoseconds after laser-target interactions~\cite{kempModelingUltrafastLaserdriven2004}. For high density or longer timescales, recombination may be important in determining the final ion charge states and the temporal evolution of ionization~\cite{townFokkerPlanckSimulationsShortpulselaser1994,wuMonteCarloApproach2017}. \cite{afshariRoleCollisionalIonization2022} discuss many recombination considerations, their potential importance in their collisional ionization simulations, and the consequences ignoring these physics may have had on their results. Regardless, a general justification for the exclusion of recombination physics is that in regions of high-density, high-temperature plasma characteristic of laser-plasma interactions, collisional effects are dominant, leading to a higher ionization rate than recombination rate~\cite{mattioliRecombinationProcessesExpansion1971}, and recombination effects only become an important consideration in regions of low temperature and density in rarified plasmas~\cite{ortnerRoleChargeTransfer2015,peterEnergyLossHeavy1991}.

\subsubsection*{Initialization of Data Table}%
\label{chap:implementation:sssec:table_init}%
Crucial to calculating our collisional ionization rates is the tabulation of numerically costly data that would be inefficient to calculate on the fly. During the initialization of the PIC simulation, we generate a data table containing the ionization cross section (\cref{chap:theory:eq:sigma}), the energy transferred to the secondary electron (\cref{chap:theory:eq:e_transf}), and the energy lost by the incident electron (\cref{chap:theory:eq:e_lost}) over a sample of incident electron energies logarithmically spaced on a linear scale. The sampled energies are calculated as follows:
\begin{equation*}
  E=e_{\text{min}}\left(\frac{e_{\text{max}}}{e_{\text{min}}}\right)^{\frac{i-1}{n_{p}-1}}\;,\; \{i\in\mathbb{N}:1\leq i\leq n_{p}\}\;,
\end{equation*}
where \(n_{p}\) is the number of points to sample the data table with and \(e_{\text{min}}/e_{\text{max}}\) are the min/max energies to sample the data table. This allows us to sample the data over large energy scales (e.g., \(\sim\)eV to \(\sim\)GeV)
that might be present within any given simulation using only a few hundred sample points. We calculate these functions using normalized kinetic energy, (\(\gamma-1\)), as opposed to another equivalent variable like the Lorentz factor, \(\gamma\), because the former better samples our function with our logarithmically spaced data points as opposed to the latter where a considerable portion of data points are located around 1. This even sampling also allows us to directly calculate the nearest index of a data entry when we later interpolate the data to continuous macro-particle energy, removing the need to do any searching through the database.

\subsubsection*{Custom Collisional Ionization Data}%
\label{chap:implementation:sssec:custom_cs}%
In addition to automatically calculating the electron-impact cross sections and related momenta values for all elements on the periodic table, the user can provide their own custom values instead. The intent behind this feature is to allow the user to calculate the ionization cross sections using a different equation than the one we chose, if they desire, without having to modify the code directly. It also allows the user to extend the applicability of this infrastructure to a wider range of potential materials and situations. %So long as users can calculate a cross section, they can ionize different chemical molecules, plastics, or materials in various states of matter. With this feature, it is possible to expand beyond electron-impact ionization and simulate proton, muon, or ion collisional ionization; we use this feature in \cref{chap:proton}.

\subsubsection*{Calculation of Collisional Ionization Rate}%
\label{chap:implementation:sssec:calc_ci_rate}%
After we populate the relevant data tables at the beginning of the simulation and reach the core PIC loop, three core aspects of the algorithm occur simultaneously every timestep: we calculate
\begin{enumerate}
  \item the ionization rate,
  \item how much momentum incident particles lose due to collisional ionization,
  \item and how much momentum we will transfer to newly generated particles.
\end{enumerate}
All three of these calculations require information from the current simulation particles, so, motivated by overall code efficiency, we handle them all concurrently as we iterate over all simulation particles once. We break down how each calculation is done.

The rate for collisional ionization for a particular ion in charge state \(i\) is $\mathsf{R}_{\text{CI},i}\equiv n_{e}\langle \overline{\sigma}_{i}v\rangle$, where %\(\mathsf{R}_{\text{CI},i}\), is
% \begin{gather}
%   \mathsf{R}_{\text{CI},i}\equiv n_{e}\langle \overline{\sigma}_{i}v\rangle\;,\;{\rm where}%
%   \label{chap:implementation:eq:collionrate}\\%
% n_{e}\langle\overline{\sigma}_{i}v\rangle=\sum_{k}\iiint_{\Omega}\sigma_{i,k}(v)vf_e(\vec{p})d^3{\vec{p}}\;.\nonumber
% \end{gather}
\begin{equation}
n_{e}\langle\overline{\sigma}_{i}v\rangle=\sum_{k}\iiint_{\Omega}\sigma_{i,k}(v)vf_e(\vec{p})d^3{\vec{p}}\;.\nonumber
\end{equation}
Here, \(n_{e}\) is the electron number density, \(v\) is the speed of the incident particle whose ionization rate we are depositing, \(\overline{\sigma}_{i}\) is the cumulative cross section for ion species \(i\) summed over its sub-shells, \(k\), and \(f_e(\vec{p})\) is the momentum distribution function of the ionizing particles.

Instead of simulating collisional ionization through particle-particle interactions, as most other PIC codes do, which requires a Monte Carlo sampling process, we exactly calculate the rate at which collisional ionization should occur given information about the macro-particles at any given timestep.

\begin{figure}[t]
\centering
\includegraphics[width=\textwidth]{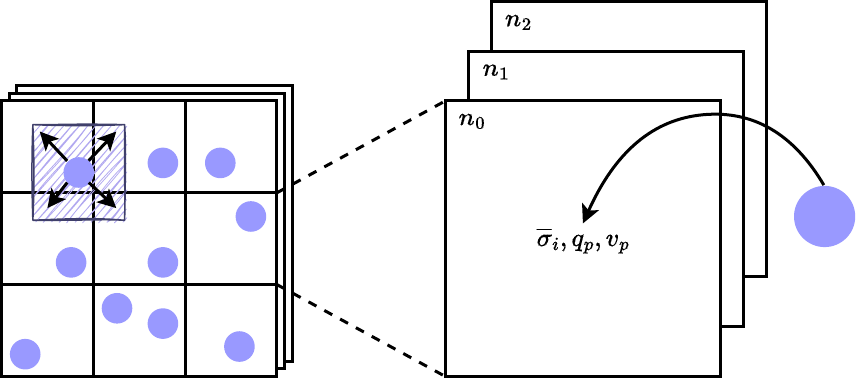}
\caption{To calculate collisional ionization rates in \textsc{Osiris}, particle information, like the particle's weight, speed, and what its cross section would be interacting with the material, is interpolated onto a series of grids. This rate information perfectly matches the corresponding grids describing the ion number densities throughout the simulation and is used to increase or decrease those densities accordingly every timestep.}
\label{chap:implementation:fig:rate_deposit}
\end{figure}

For each neutral species, a grid is first created for each ionization state with the same dimensions as the grid used for the electric field. We iterate over every particle in each collisionally ionizing species that the user specifies and deposit their contribution to the total ionization rate onto the grid. In practice, to evaluate the collision rate and deposit the ionization rate, we iterate over every particle and deposit their particle weight, velocity, and cross section onto the ionization rate grid for each ionization level and distribute this quantity according to the particle's shape function. The total ionization rate, \({\mathsf{R}_{\text{CI},i}}_{j}\), for ion species \(i\) is therefore calculated at time step \(m\) through the summation over \(P\) particles
\begin{equation}
  {\mathsf{R}^{m}_{\text{CI},i}}_{j}=\sum_{p=1}^{P}S\left(\vec{X}_{j}-\vec{x}_{p}\right)|q_{p}|\overline{\sigma}_{i}\left(T_{p}\right)|\vec{v}_{p}|\;,%
  \label{chap:implementation:eq:discretized_ci_rate}%
\end{equation}
where \(S\left(\vec{X}_{j}-\vec{x}_{p}\right)\) is the particle shape factor at grid cell \(j\) with position \(\vec{X}_{j}\) and particle position \(\vec{x}_{p}\), \(|q_{p}|\) is the particle's weight (or equivalently the absolute value of charge in \textsc{Osiris}), \(\overline{\sigma}_{i}(T_{p})\) is the cross section of the ion species for an electron with energy \(T_{p}\), and \(|\vec{v}_{p}|\) is the speed of the incident particle.

As this rate is calculated, we remove energy from the incident particles contributing to the ionization rate and calculate the amount of energy being transferred to newly created particles. % Similar to the \textsc{Smilei}~\cite{derouillatSmileiCollaborativeOpensource2018} code,  the change in momentum is calculated in the following . Since the ionizable ions are stationary within our infrastructure, we inherently perform these calculations in the ion rest frame, making some of our expressions simpler than their counterparts presented by \textsc{Smilei}.
As we iterate over every particle to calculate the ionization rate, we actively keep track of the change in all components of that particle's momentum, \(\vec{p}\), as it causes ionization with each present ion state for a given neutral. Noting that for the mobile ion model we take the approximation that the ions are slow, the change in momentum of a single particle with label \(p\) is calculated as
\begin{equation}
  \label{chap:implementation:eq:momlost}%
\Delta\vec{p}_{p}=\sum_{i,j}S\left(\vec{X}_{j}-\vec{x}_{p}\right)\overline{\sigma}_{i}(T_{p})|\vec{v}_{p}|{n_{i}}_{j}\left(\alpha_{\varepsilon,i}-1\right)\vec{p}_{p}\Delta t\;,\;
\end{equation}
where
\begin{equation}
\alpha_{\varepsilon,i}=\frac{\sqrt{(\gamma_{p}-\overline{\varepsilon_{i}})^{2}-1}}{\sqrt{\gamma_{p}^{2}-1}}\;.\label{eqn:alpha_e}
\end{equation}
\({n_{i}}_{j}\) is the local ion density from \cref{chap:implementation:eq:density_profile} in grid cell \(j\) for charge state $i$ and $\overline{\varepsilon_{i}}$ is the average energy loss of the incident particle in units of $m_ec^2$. It also needs to be ensured that this scheme is numerically stable and that we do not remove more momenta from the particle than it started with.

If the rate of momentum lost is given by \(\mathsf{R}_{\varepsilon}=\overline{\sigma}_{i}(T_{p})|\vec{v}_{p}|{n_{i}}_{j}\left(1-\alpha_{\varepsilon,i}\right)\),
% \footnote{Yes, this expression is written correctly and has an added negative sign versus its counterpart just above. This was done to make several of the following expressions easier to write. The verification of equality between \cref{chap:implementation:eq:momlost,chap:implementation:eq:momlost_fd} can be left as an exercise to the reader with the knowledge that \(\mathsf{R}_{\varepsilon}\) is always positive given that \(0\leq\alpha_{\varepsilon}\leq1\).}
then the semi-implicit, center-differenced (Runge-Kutta) expansion of this term can be written as
\begin{equation}
\mathsf{R}_{\varepsilon,\text{RK}}=\frac{\mathsf{R}_{\varepsilon}\Delta t}{1+\frac{1}{2}\mathsf{R}_{\varepsilon}\Delta t}\;.
\end{equation}
For large $\mathsf{R}_{\varepsilon}\Delta t$ (in either the positive or negative infinity limit), $\mathsf{R}_{\varepsilon,\text{RK}}$ converges to 2 while we desire to clamp this to a maximum value of 1. The physical justification for this upper limit is such: an incident electron causing ionization has, at most, \(\vec{p}_{p}\) amount of momentum to lose in a given timestep. In other words, a value of \(\mathsf{R}_{\varepsilon,\text{RK}}=1\) means that the incident electron has lost all of its momentum in one go, and a value of 2 would imply that an electron has lost twice its original momentum. With this constraint, \cref{chap:implementation:eq:momlost} can be rewritten as
\begin{equation}
  \Delta\vec{p}_{p}=-\sum_{i,j}\min\left(\mathsf{R}_{\varepsilon,\text{RK}}, 1\right)S\left(\vec{X}_{j}-\vec{x}_{p}\right)\vec{p}_{p}\;.%
  \label{chap:implementation:eq:momlost_fd}%
\end{equation}%

Finally, \cref{chap:implementation:eq:momlost_fd} does not guarantee that \(\Delta\vec{p}_{p}\leq\vec{p}_{p}\) over the sum of all \(i\) ionization states. It is unlikely for this to occur, but simulation parameters can be altered such that this condition is violated. So, once we finish calculating that particle's contribution to the ionization rate, we update its momentum via
\begin{equation}
  \vec{p}_{p}'=\vec{p}_{p}+\text{sign}\left(\Delta\vec{p}_{p}\right)\min\left(|\Delta\vec{p}_{p}|, |\vec{p}_{p}|\right)\;.%
  \label{chap:implementation:eq:momlost_update}%
\end{equation}

It should be noted that within this scheme, there is no inherent scattering of the particles included in this calculation that would change \(\hat{\vec{p}}_{p}\). This is an inaccuracy but not a substantial one as experiments measuring the fast electrons (the electrons having done the ionization) leaving the material show that they mostly leave collinear with their incident momentum~\cite{ehrhardtCollisionalIonizationHelium1972}. (It should be noted, though, that there are few experiments reporting these measurements, and the few that do are for specific, low-Z materials with low-energy (read classical regime) incident electrons.) Additionally, momentum is removed from all current simulation particles participating in ionization regardless of whether a new particle is generated that timestep.

\subsubsection*{Immobile Ions}%
\label{chap:implementation:sssec:ci_rate_immobile}%
\begin{figure}[p]
  \centering
  % \begin{subfigure}[c]{0.37\textwidth}
  % \centering
  % \includegraphics[width=\textwidth]{figs/pic_cell_1d_w_markers.pdf}
  % \label{chap:implementation:fig:pic_cell_1d}
  % \end{subfigure}
  % \hfill
  % \begin{subfigure}[c]{0.54\textwidth}
  % \centering
  % \includegraphics[width=\textwidth]{figs/pic_cell_2d_w_markers.pdf}
  % \label{chap:implementation:fig:pic_cell_2d}
  % \end{subfigure}
  % \\
  % \begin{subfigure}[c]{0.54\textwidth}
  % \centering
  % \includegraphics[width=\textwidth]{figs/pic_cell_3d_w_markers.pdf}
  % \label{chap:implementation:fig:pic_cell_3d}
  % \end{subfigure}
  \includegraphics[width=\textwidth]{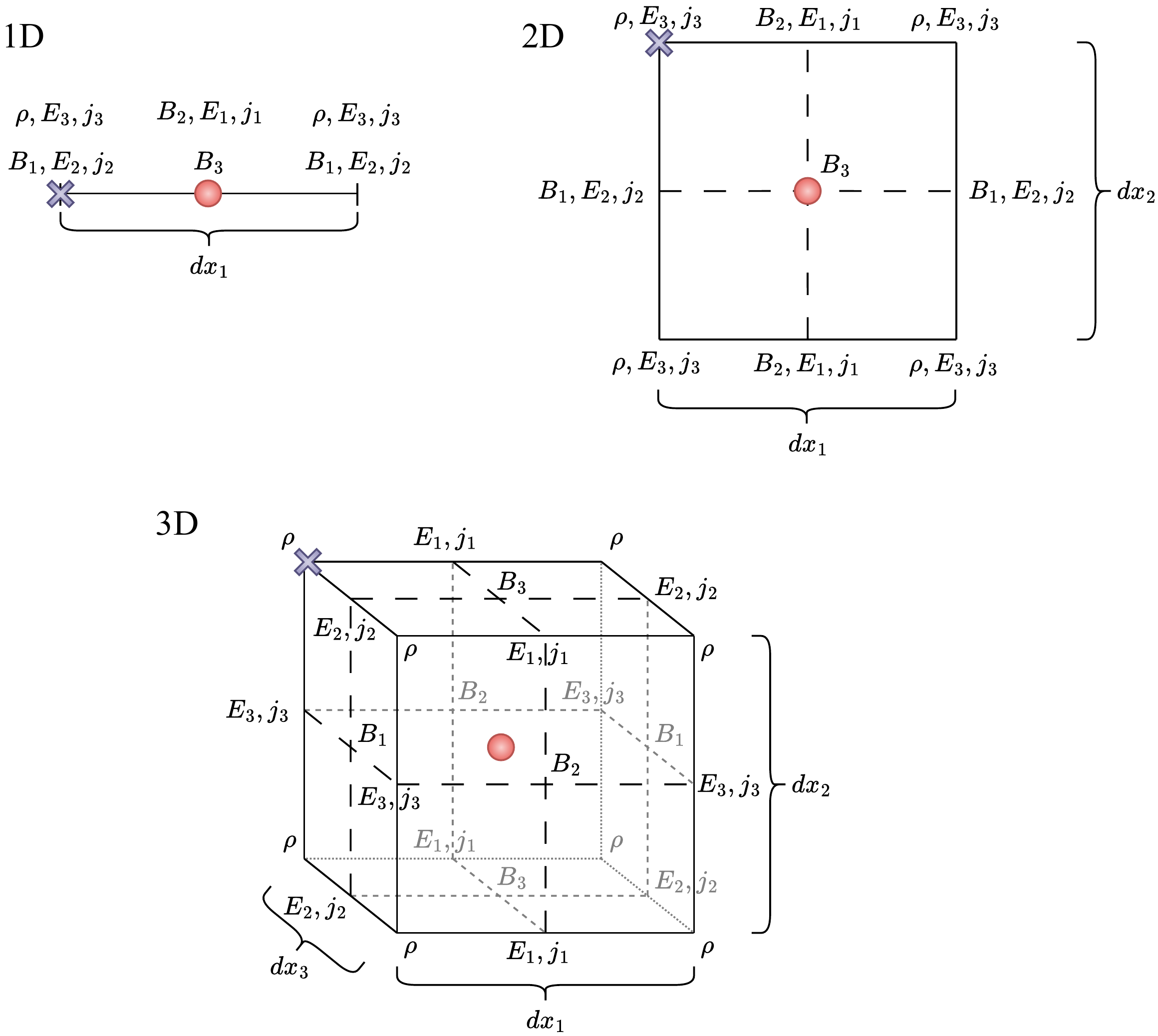}
  \caption{The structure of a typical PIC cell in 1-, 2-, and 3-dimensions highlighting where Yee-staggered quantities, like the components of the electric and magnetic field, are stored relative to the grid. The purple `X' indicates the cell corner where density information is calculated and subsequently where, by default, our collisional rates are calculated. For even-order shape functions, there is no issue as the field ionization rates are also averaged onto each cell corner. For odd-order shape functions, however, the field and collisional ionization rates are averaged to the center of the cell volume, indicated with the red circle.}
  \label{chap:implementation:fig:pic_cell_w_markers}
\end{figure}

At this point, the approach to handling immobile and mobile ions diverges in implementation. Recall from \cref{chap:implementation:sssec:fi_immobile} that depending on the order of shape function used, where we calculate the local electric field with respect to a grid cell varies. Thus, careful attention goes to ensure that the new collisional ionization rates we produce are calculated at those same positions. By default, the collisional ionization rates from \cref{chap:implementation:eq:discretized_ci_rate} are centered at the cell corners with the density values, \(|q_{p}|\). For an odd shape function, however, a level of re-interpolation is conducted to find the average ionization rate at the center of the cell volume (see \cref{chap:implementation:fig:pic_cell_w_markers}). These averages are found via
\begin{align*}
  &\mathsf{R}=\frac{1}{2}\left(\mathsf{R}_{j_{1}} + \mathsf{R}_{j_{1}+1}\right) &\text{for 1D,\phantom{ and}}\\
  &\mathsf{R}=\frac{1}{4}\left(\mathsf{R}_{j_{1},j_{2}} + \mathsf{R}_{j_{1}+1,j_{2}} + \mathsf{R}_{j_{1},j_{2}+1} + \mathsf{R}_{j_{1}+1,j_{2}+1}\right) &\text{for 2D, and}\\
&\begin{aligned}
  \mathsf{R}=\frac{1}{8}\left(\vphantom{\mathsf{R}_{j_{1}+1,j_{2}+1,j_{3}+1}}\right. &\mathsf{R}_{j_{1},j_{2},j_{3}} \\
  & + \mathsf{R}_{j_{1}+1,j_{2},j_{3}} + \mathsf{R}_{j_{1},j_{2}+1,j_{3}} + \mathsf{R}_{j_{1},j_{2},j_{3}+1}\\
  & + \mathsf{R}_{j_{1}+1,j_{2}+1,j_{3}} + \mathsf{R}_{j_{1}+1,j_{2},j_{3}+1} + \mathsf{R}_{j_{1},j_{2}+1,j_{3}+1}\\
  & \left.+ \mathsf{R}_{j_{1}+1,j_{2}+1,j_{3}+1}\right)
\end{aligned}& \text{for 3D,\phantom{ and}}
\end{align*}
where \(j_{d}\) refers to the cell position in the \(d^{\text{th}}\) dimension.

Finally, it is convenient for us to calculate the amount of momentum transferred to a newly ionized particle during this loop. We create another aforementioned grid that collects all the momentum contributions, \(\vec{p}_{w}\), from each incident ionizing electron. The amount of momenta deposited in a given cell, \(j\), is calculated via
\begin{equation}
  \label{chap:implementation:eq:momtransf_immobile}%
  n_{\text{ionized}}{\left\langle\vec{p}_{w}\right\rangle}_{j}=\sum_{p=1}^{P}\sum_{i}S\left(\vec{X}_{j}-\vec{x}_{p}\right)|q_{p}|\overline{\sigma}_{i}(T_{p})|\vec{v}_{p}|{n_{i}}_{j}\alpha_{w,i}\vec{p}_{p}\Delta t\;,\;
  \end{equation}
  where
  \begin{equation}
\alpha_{w,i}=\frac{|\vec{p}_{w}|}{|\vec{p}_{p}|}=\frac{\sqrt{\overline{w_{i}}(2+\overline{w_{i}})}}{\sqrt{\gamma_{p}^{2}-1}}\;.\label{eqn:alphaw}
  \end{equation}
\(n_{\text{ionized}}\) is the number density of the ionized electrons, which is expanded on in \cref{chap:implementation:sssec:immobile_inject}.

As a reminder, we define the rate of ionization per ion as \(\mathsf{R}=|q_{p}|\overline{\sigma}_{i}(T_{p})|\vec{v}_{p}|\). As \cref{chap:implementation:eq:momtransf_immobile} is currently written, if \(\mathsf{R}\Delta t>1\), or the regime where the complete ionization of one or more states occurs within a single timestep, then we quickly begin to overestimate the actual amount of momentum transferred by the incident particles. Additionally, suppose the number of ionized particles is small compared to the maximum ionization state (\(P\lessapprox z_{\text{max}}\)). In that case, it becomes increasingly difficult for those few particles to capture all of the relevant physics. These issues can be partially rectified with the following adjustments. The first is using a Runge-Kutta scheme for the ionization rate.
\begin{equation*}
  \mathsf{R}_{\text{RK}}=\frac{\mathsf{R}\Delta t}{1+\frac{1}{2}\mathsf{R}\Delta t}
\end{equation*}%
This finite differencing is sufficient in handling extreme ranges of \(\mathsf{R}\Delta t\) and provides the calculation with an appropriate amount of stability. The second is a switch to ensure that we do not over-prescribe an amount of momentum transferred when we fully ionize particular levels. Using the same logic as we did with the momentum lost by incident particles, if \(\mathsf{R}_{\text{RK}}\geq1\), or the ion becomes completely ionized, then the amount of momentum transferred by an incident particle is \({n_{i}}_{j}\alpha_{w,i}\vec{p}_{p}\). If not, then the amount of momentum deposited is proportional to how much of the ion density gets ionized, \(\mathsf{R}_{\text{RK}}\), multiplied by the amount of momentum transferred.
\begin{equation}
  n_{\text{ionized}}{\left\langle\vec{p}_{w}\right\rangle}_{j}=\sum_{p=1}^{P}\sum_{i}
  \begin{cases}
    S\left(\vec{X}_{j}-\vec{x}_{p}\right){n_{i}}_{j}\alpha_{w,i}\vec{p}_{p} & \mathsf{R}_{\text{RK}}\geq1\\
    \mathsf{C}_{\text{inj}}\mathsf{R}_{\text{RK}}S\left(\vec{X}_{j}-\vec{x}_{p}\right){n_{i}}_{j}\alpha_{w,i}\vec{p}_{p} & \mathsf{R}_{\text{RK}}<1
  \end{cases}%
  \label{chap:implementation:eq:center_diff_momtransf_immobile}%
\end{equation}
The constant \(\mathsf{C}_{\text{inj}}\) is a term we include solely as a consequence of how we determine when to inject new macro-particles. Defined as
\begin{equation*}
  \mathsf{C}_{\text{inj}} = \frac{1}{1-\frac{1}{2n_{\text{ppc}}}}\;,
\end{equation*}
with \(n_{\text{ppc}}\) being the number of particles per cell specified for the ionized electron species, the coefficient acts as a correction to compensate for the difference between the charge assigned to the macro-particle at injection and the actual amount of charge ionized.

The data in this momentum grid is never reset but accumulates from timestep to timestep. Only once a new particle is injected due to ionization do we use and adjust the values in this grid; this is discussed in \cref{chap:implementation:sssec:immobile_inject}.

\subsubsection*{Mobile Ions}%
\label{chap:implementation:sssec:ci_rate_mobile}%
Recall that much of the relevant quantities we previously calculated are now stored with each ion macro-particle instead of on a grid. This especially becomes an important distinction when calculating the amount of momentum we transfer to newly ionized electron macro-particles. Instead of having a grid that stores the cumulative amount of momentum transferred to newly ionized electrons from all of the ionizing interactions, as with immobile ions, each ion macro-particle now collects an amount of momentum that will be transferred to the new electron it generates upon promotion. The fact that one electron is generated per ionization level per macro-particle simplifies this process.

We still abstract away any direct collision between the incident, ionizing electrons and the ion macro-particles themselves, so we create \(z_{\text{max}}\) momentum transfer grids, one for each ionization state of our ion species, that collect the cumulative momentum transferred from each of the ionizing interactions. In general, the amount of momentum transferred per ion is calculated via
\begin{equation}
  {\left\langle\vec{p}_{w}\right\rangle}_{i,j}=\sum_{p=1}^{P}S\left(\vec{X}_{j}-\vec{x}_{p}\right)|q_{p}|\overline{\sigma}_{i}(T_{p})|\vec{v}_{p}|\alpha_{w,i}\vec{p}_{p}\Delta t\;.%
  \label{chap:implementation:eq:momtransf_mobile}%
\end{equation}
As with the previous calculations, some additional work is done to ensure that our calculation is numerically stable and does not overprescribe the amount of momentum transferred. Again, using the Runge-Kutta scheme for the ionization rate, we can rewrite the above formula in the following manner:
\begin{equation}
  {\left\langle\vec{p}_{w}\right\rangle}_{i,j}=\sum_{p=1}^{P}\min\left(\mathsf{R}_{\text{RK}},1\right)S\left(\vec{X}_{j}-\vec{x}_{p}\right)\alpha_{w,i}\vec{p}_{p}\;.%
  \label{chap:implementation:eq:center_diff_momtransf_mobile}%
\end{equation}

Once we have finished iterating over all of the ionizing particles and calculated all of our momentum transfer grids, we then interpolate values from those grids onto each ion macro-particle. From one timestep, \(m\), to the next, \(m+1\), each ion macro-particle's stored momentum to transfer, \(\vec{p}_{\text{tr}}\), is updated via
\begin{equation}
  \vec{p}_{\text{tr}}^{m+1}=\vec{p}_{\text{tr}}^{m}+\zeta\sum_{j}S\left(\vec{X}_{j}-\vec{x}_{p}\right){\left\langle\vec{p}_{w}\right\rangle}_{i,j}\;.%
  \label{chap:implementation:eq:momtransf_mobile_update}%
\end{equation}
\(\zeta\) is the measure of how ionized the macro-particle is currently and was previously defined with \cref{chap:implementation:eq:zeta}. Only the ion macro-particles belonging to the species representing the \(i^{\text{th}}\) charge state draws momentum from the \(i^{\text{th}}\) momentum transfer grid. This accumulated momentum is discussed further in \cref{chap:implementation:sssec:mobile_inject}.

\subsubsection*{Interpolation of Relevant Quantities}%
\label{chap:implementation:sssec:interpolation}%
Regardless of ion mobility, the values we use for the cross section, energy lost, and energy transferred are crucial to these calculations. Recall that we populate a data table with these calculated values for an ion species at the program's initialization. We make use of this data table by interpolating and extrapolating values from this table rather than explicitly evaluating \cref{chap:theory:eq:sigma,chap:theory:eq:e_transf,chap:theory:eq:e_lost} for every particle at every timestep. The kinetic energy, \(T\), of an incident electron is calculated, and depending on whether the energy falls within the limits of \(e_{\text{min}}\) and \(e_{\text{max}}\), we either interpolate between values already in the data table or extrapolate beyond them. The same formula can be used in either case. An interpolated cross-section value, \(\overline{\sigma}^{\prime}\), for example, is calculated via
\begin{gather*}
  \overline{\sigma}^{\prime}=\left(\overline{\sigma}_{j+1}-\overline{\sigma}_{j}\right)\delta+\overline{\sigma}_{j}\;,\;{\rm where}\\
  \delta=x-j\;.
\end{gather*}
\(\overline{\sigma}_{j}\) and \(\overline{\sigma}_{j+1}\) are values from the cross-section table at indices \(j\) and \(j+1\), respectfully. Due to the linear spacing of energies, the index of the value closest to the particle's kinetic energy may be calculated efficiently by setting \(j=\lfloor x\rfloor\), where \(x\) is determined via\footnote{We write these formulas assuming the programming language of choice uses one-based indexed arrays, as in Fortran.}
\begin{equation*}
  x=\log_{10}\left(T/e_{\text{min}}\right)\frac{n_{p}-1}{\log_{10}\left(e_{\text{max}}/e_{\text{min}}\right)}+1\;.
\end{equation*}
If the kinetic energy of the particle is below \(e_{\text{min}}\), then \(j\) is automatically set to 1, and a value is extrapolated using \(\overline{\sigma}_{1}\) and \(\overline{\sigma}_{2}\). Similarly, if the particle's energy is above \(e_{\text{max}}\), then \(j\) is automatically set to \(n_{p}-1\) and a value extrapolated using \(\overline{\sigma}_{n_{p}-1}\) and \(\overline{\sigma}_{n_{p}}\).

\subsection{Ion Density Advancement}%
\label{chap:implementation:ssec:dens_adv}%
Once the cumulative ionization rate has been calculated, we can deterministically advance the densities of each ionization state of the neutral object. We aim at this point to evaluate \cref{chap:implementation:eq:dni_dt} given a rate of ionization, \(\mathsf{R}\), and the state densities, \(n_{i}\). A series of differential equations describing this evolution can be written as follows:
% {%
% \begin{align}%
% &\begin{aligned}
%   \hphantom{\frac{dn_{z_{\text{max}}-1}}{dt}\;}%
%   {\frac{dn_{i}}{dt}}&=n_{i-1}\mathsf{R}_{i-1}-n_{i}\mathsf{R}_{i}\\%
%   &\setbox0\hbox{=}\mathrel{\makebox[\wd0]{\hfil\(\Downarrow\)\hfil}}% center the arrow
% \end{aligned}\nonumber \\%
% &%
% \begin{rparent}%
%   \dv{n_{0}}{t}&=-n_{0}\mathsf{R}_{0}\\%
%   \dv{n_{1}}{t}&=n_{0}\mathsf{R}_{0}-n_{1}\mathsf{R}_{1}\\%
%   \dv{n_{2}}{t}&=n_{1}\mathsf{R}_{1}-n_{2}\mathsf{R}_{2}\\%
%   &\setbox0\hbox{=}\mathrel{\makebox[\wd0]{\hfil\vdots\hfil}}\\% center the vdots
%   \dv{n_{z_{\text{max}}-1}}{t}&=n_{z_{\text{max}}-2}\mathsf{R}_{z_{\text{max}}-2}-n_{z_{\text{max}}-1}\mathsf{R}_{z_{\text{max}}-1}\\%
%   \dv{n_{z_{\text{max}}}}{t}&=n_{z_{\text{max}}-1}\mathsf{R}_{z_{\text{max}}-1}%
% \end{rparent}%
% \label{chap:implementation:eq:bateman_eqs}%
% \end{align}%
% }%
{%
\begin{align}%
  \frac{dn_{i}}{dt}&=n_{i-1}\mathsf{R}_{i-1}-n_{i}\mathsf{R}_{i}\;,\nonumber\\%
  &\setbox0\hbox{=}\mathrel{\makebox[\wd0]{\hfil\(\Downarrow\)\hfil}}\nonumber\\% center the arrow
  \frac{dn_{0}}{dt}&=-n_{0}\mathsf{R}_{0}\;,\nonumber\\
  \frac{dn_{1}}{dt}&=n_{0}\mathsf{R}_{0}-n_{1}\mathsf{R}_{1}\;,\label{chap:implementation:eq:bateman_eqs}\\%
   \frac{dn_{2}}{dt}&=n_{1}\mathsf{R}_{1}-n_{2}\mathsf{R}_{2}\;,\nonumber\\%
   &\setbox0\hbox{=}\mathrel{\makebox[\wd0]{\hfil\vdots\hfil}}\;,\nonumber\\% center the vdots
   \frac{dn_{z_{\text{max}}-1}}{dt}&=n_{z_{\text{max}}-2}\mathsf{R}_{z_{\text{max}}-2}-n_{z_{\text{max}}-1}\mathsf{R}_{z_{\text{max}}-1}\;,\nonumber\\%
  \frac{dn_{z_{\text{max}}}}{dt}&=n_{z_{\text{max}}-1}\mathsf{R}_{z_{\text{max}}-1}\;.\nonumber
\end{align}%
}%

In addition to this set of equations describing the evolution of the ion's number densities, the increase in the electron number density can be written as follows
\begin{equation}
  \frac{dn_{e}}{dt}=n_{0}\mathsf{R}_{0}+n_{1}\mathsf{R}_{1}+\cdots+n_{z_{\text{max}}-1}\mathsf{R}_{z_{\text{max}}-1}\;.%
  \label{chap:implementation:eq:electron_den_advance}%
\end{equation}
Again, we only allow the electron number density to increase because we ignore recombination effects.
\subsubsection{Immobile Ions}%
\label{chap:implementation:sssec:advance_immobile}%
In the case of immobile ions, two approaches could be used to solve this set of differential equations. The first is a semi-implicit, center-differenced scheme to take us from the \(m^{\text{th}}\) timestep to the next. Taking the differential equation corresponding to the neutral, to begin with, we can write out the following:
\begin{align*}
  \frac{dn_{0}}{dt}&=-n_{0}\mathsf{R}_{0}\\%
  \frac{n_{0}^{m+1}-n_{0}^{m}}{\Delta t}&=-n_{0}^{m+1/2}\mathsf{R}_{0}\\%
  &=-\frac{1}{2}\left(n_{0}^{m}+n_{0}^{m+1}\right)\mathsf{R}_{0}\\%
  n_{0}^{m+1}\left(1+\frac{\mathsf{R}_{0}\Delta t}{2}\right)&=n_{0}^{m}\left(1-\frac{\mathsf{R}_{0}\Delta t}{2}\right)\\%
  \Delta n_{0}=n_{0}^{m+1}-n_{0}^{m}&=\frac{-\mathsf{R}_{0}\Delta t}{1+\frac{1}{2}\mathsf{R}_{0}\Delta t}n_{0}^{m}\;.%
\end{align*}
As with our previous differencing attempts, some additional checks are done to ensure that the coefficient before \(n_{0}^{m}\) does not exceed a value of 1. From this, a generalized scheme can be extended to solve the differential equation for any ionization state:
\begin{align*}
  \Delta n_{i}=n_{i}^{m+1}-n_{i}^{m}&=\frac{\mathsf{R}_{i-1}\Delta t}{1+\frac{1}{2}\mathsf{R}_{i-1}\Delta t}n_{i-1}^{m}-\frac{\mathsf{R}_{i}\Delta t}{1+\frac{1}{2}\mathsf{R}_{i}\Delta t}n_{i}^{m}\;.%
\end{align*}
This structure guarantees overall charge conservation, \(\sum_{i}(n^{m+1}_{i}-n^{m}_{i})=0\), while still providing second-order accuracy. Given the internal structure of the neutral object, discussed in length above, we do not require the calculation of \(n_{i}\), and we capture the same information using the \(\overline{Z}_{i}\) grids:
\begin{align}
\Delta\overline{Z}_{0}=\overline{Z}_{0}^{m+1}-\overline{Z}_{0}^{m}&=-\frac{\mathsf{R}_{0}\Delta t}{1+\frac{1}{2}\mathsf{R}_{0}\Delta t}\overline{Z}_{0}^{m}\;,\\%
\Delta\overline{Z}_{i}=\overline{Z}_{i}^{m+1}-\overline{Z}_{i}^{m}&=\frac{\mathsf{R}_{i-1}\Delta t}{1+\frac{1}{2}\mathsf{R}_{i-1}\Delta t}\overline{Z}_{i-1}^{m}-\frac{\mathsf{R}_{i}\Delta t}{1+\frac{1}{2}\mathsf{R}_{i}\Delta t}\overline{Z}_{i}^{m}\;,\;\{0<i<\text{z}_{\text{max}}\}\;,\\%
\Delta\overline{Z}_{\text{z}_{\text{max}}}=\overline{Z}_{\text{z}_{\text{max}}}^{m+1}-\overline{Z}_{\text{z}_{\text{max}}}^{m}&=\frac{\mathsf{R}_{\text{z}_{\text{max}}-1}\Delta t}{1+\frac{1}{2}\mathsf{R}_{\text{z}_{\text{max}}-1}\Delta t}\overline{Z}_{\text{z}_{\text{max}}-1}^{m}\;.%
  \label{chap:implementation:eq:denadvance}%
\end{align}

\subsubsection{Mobile Ions}%
\label{chap:implementation:sssec:advance_mobile}%
We still aim to solve \cref{chap:implementation:eq:bateman_eqs,chap:implementation:eq:electron_den_advance} to adjust the number density of each ion state with mobile, ionizable macro-particles. Recall that mobile ions carry a quantity, \(\zeta\), which measures how far along that macro-particle has been ionized from one state to its next.

Making use of the ionization rate grids we calculated previously, a rate, with all the contributions from field and collisional ionization, is interpolated onto each ion macro-particle via
\begin{equation}
  \mathsf{R}_{i}=\sum_{j}S\left(\vec{X}_{j}-\vec{x}_{p}\right)\mathsf{R}_{\text{CI},i}+\mathsf{R}_{\text{FI},i}\left(S\left(\vec{X}_{j}-\vec{x}_{p}\right)E_{j}\right)\;.
\end{equation}

Using this rate, we update each ion's \(\zeta\) value. Using the all-to-familiar approach, we rewrite the ionization rate in the following form
\begin{equation*}
  \mathsf{R}_{\text{RK}}=\frac{\mathsf{R}_{i}\Delta t}{1+\frac{1}{2}\mathsf{R}_{i}\Delta t}\;.
\end{equation*}
Then, from one timestep to the next, each particle's \(\zeta\) advances via
\begin{equation}
  \zeta^{m+1}_{i}=\zeta^{m}_{i}-\min(\mathsf{R}_{\text{RK}},1)\zeta^{m}_{i}\;.%
  \label{chap:implementation:eq:zeta_advance}%
\end{equation}

More generally, each particle's change in \(\zeta\) is equivalent to the equations describing the change in ion number density (\cref{chap:implementation:eq:bateman_eqs}), and we can write this change as the following differential equation. 
\begin{equation}
  % \dv{\zeta_{i}}{t}&=\cancelto{0}{\zeta_{i-1}\mathsf{R}_{i-1}}-\zeta_{i}\mathsf{R}_{i}%
  \frac{d\zeta_{i}}{dt}=\zeta_{i-1}\mathsf{R}_{i-1}-\zeta_{i}\mathsf{R}_{i}%
  \label{chap:implementation:eq:movion_rate}%
\end{equation}
A comparison between \cref{chap:implementation:eq:zeta_advance} and \cref{chap:implementation:eq:movion_rate} shows that the actual implementation we propose does not include the source term in the latter equation. However, a closer look at the overall algorithm reveals that we properly captured this differential equation's behavior.

As it is currently implemented, the value of \(\zeta\) only ever decreases as a macro-particle becomes more ``ionized.'' Even if macro-particles representing a lower charge state see a decrease in their \(\zeta\) value, macro-particles in higher charge states never increase their \(\zeta\). So, while we handle the loss term in this portion of the algorithm, the source term is dealt with by promoting ion macro-particles from lower charge states to higher states. Despite never increasing a particle's \(\zeta\), this implementation proves sufficient to properly model the solutions of the differential equation we wrote above. To relate this discussion to modeling \cref{chap:implementation:eq:bateman_eqs,chap:implementation:eq:electron_den_advance}, the changes in the ion number densities are handled entirely via this particle creation and deletion process as \(\zeta\) values decrease and macro-particles get promoted.

\subsection{Ionized Particle Injection}%
\label{chap:implementation:ssec:part_inject}%
Until this point, we have calculated the amount of ionization that has occurred throughout the simulation but not produced any new macro-particles. This is the final step of the algorithm. The significant difference between particle injection in the case of immobile and mobile ions is how we determine when particles get injected and what momenta to assign them. We describe those key differences below. It is worth noting that in simulations including only field ionization, newly created particles are injected with zero momentum for both mobile and immobile ions~\cite{beckerPlateauAbovethresholdIonization2018}. %The rest of the following discussion beyond how momentum is handled is the same. However, if a simulation is run with collisional ionization, the following discussion, particularly the momentum calculations, is adhered to exactly as written, regardless of whether field ionization is included in the simulated physics.

\subsubsection{Immobile Ions}%
\label{chap:implementation:sssec:immobile_inject}
After advancing the densities in each \(\overline{Z}_{i}\) grid, particle injection is determined and carried out cell-by-cell. For each cell, we compare how much the charge density has changed from the last timestep to the current. If the cumulative charge generated from ionizing each state is sufficient, we create and inject new macro-particles into the current grid cell. The number of particles to inject into a cell is determined via
\begin{equation}
  n_{\text{inj}}=\left\lfloor \frac{\overline{Z}^{*}_{m}}{z_{\text{max}}}n_{\text{ppc}}+0.5\right\rfloor-\left\lfloor \frac{\overline{Z}^{*}_{m-1}}{z_{\text{max}}}n_{\text{ppc}}+0.5\right\rfloor\;,%
  \label{chap:implementation:eq:ninj}%
\end{equation}
where \(\overline{Z}^{*}_{m}\) is the quantity from \cref{chap:implementation:eq:total_charge} at timestep \(m\), and \(n_{\text{ppc}}\) is the number of particles per cell the user specifies for the species representing the ionized electrons. Particles can either be injected at random locations within the cell or injected collinearly and evenly spaced along the center lines of each cell volume. This is included to address some noise issues. For every particle injected into a cell, \(j\), we give them a weight of
\begin{equation}
  q_{\text{inj}}=-\frac{z_{\text{max}}\overline{Z}_{\textsc{Prof},j}}{n_{\text{ppc}}}\;.
\end{equation}
It is essential to highlight that the full number of electrons available, \(n_{\text{ppc}}\), are injected into a cell only when that cell becomes fully ionized to \(z_{\text{max}}\). For example, if the neutral species has a \(z_{\text{max}}\) of 3 and the user sets \(n_{\text{ppc}}\) to be 9, then only three ionized electron macro-particles get injected per ionization state. To be explicit, \(n_{\text{ppc}}/z_{\text{max}}\) is the number of macro-particles representing the ionized electrons from each ionization level of the neutral for a specific grid cell.

At the point of injection, each new electron macro-particle is assigned momenta using the grid we calculated before in \cref{chap:implementation:sssec:ci_rate_immobile} via \cref{chap:implementation:eq:center_diff_momtransf_immobile}. Recall that each cell in this momentum grid stored the cumulative momentum that would have been transferred to ionized electrons from every ionization event. When one or more particles are injected into this cell, we take all the momenta stored in the given grid cell, \(j\), and distribute it evenly amongst the newly created particles. Each new particle receives a momenta prescribed by
\begin{equation}
  \vec{p}_{\text{new}}=\frac{n_{\text{ionized}}{\left\langle\vec{p}_{w}\right\rangle}_{j}}{n_{\text{inj}}|q_{\text{inj}}|}\;.
\end{equation}
After distributing this momentum, the value in this cell of the momenta grid is reset back to zero so more momenta can accumulate starting with the next time step.

\subsubsection{Mobile Ions}%
\label{chap:implementation:sssec:mobile_inject}
After we update each macro-particle's \(\zeta\) value, we check to see if the condition \(\zeta<r\) is satisfied, where \(r\) is that particle's promotion threshold. If a particle meets this criterion, it is marked for promotion, and we continue updating the rest of the macro-particles for the other charge states.

Once we have iterated over every particle in all charge states, the following actions are set into motion for the marked particles. A new ion macro-particle is created for the species in the successive charge state. This new particle has the same position and momentum as the current macro-particle we are promoting. But, we increase the charge (or weight) of the new particle by a factor of \((i+1)/i\), where \(i\) is the charge state of the species the particle is being promoted from. (This factor is excluded if a particle is promoted from a charge state of zero to +1.)

In addition to creating a higher charge-state ion, a new electron macro-particle is injected into the simulation. We assign this new electron a weight of \(-q_{i}/i\), where \(q_{i}\) is the weight of the ion macro-particle being promoted. (Again, if it is an ion with a charge state of 0 being ionized, the electron is just assigned a weight of \(-q_{i}\).) The new electron is still created within the same cell as its parent ion but assigned a randomized position. This is to prevent the electron and ion particles from being created at the same location where their current contributions would cancel each other out, which essentially removes their contribution to any of the plasma dynamics~\cite{mayEnhancedStoppingMacroparticles2014,mayAccelerationTransportElectrons2017}. (This is less of an issue at higher ionization stages where the charges of either particle do not exactly cancel out but is an issue for the first ionization state.)

Finally, using the momenta accumulated via \cref{chap:implementation:eq:momtransf_mobile_update} by the parent macro-particle, the electron is assigned a momentum calculated via
\begin{equation}
  \vec{p}_{e}=\frac{\vec{p}_{\text{tr},i}}{1-\zeta_{i}}\;.
\end{equation}
The factor of \((1-\zeta_{i})^{-1}\) is an adjustment to account for the fact that the original ion macro-particle may not have been completely ionized before undergoing a promotion. To avoid prescribing too little momenta to the ionized electron, we scale the momentum the ion has accumulated by a factor equivalent to how much it had left to be ionized. Once we have finished creating all the new ion and electron macro-particles, the original parent ion that underwent promotion is deleted from the simulation. The new ion macro-particle is assigned a value \(\zeta=1\), a new random promotion threshold, and starts with \(\vec{p}_{\text{tr}}=0\), and the process repeats.

\section{Code Verification}%
\label{chap:implementation:sec:pop}%
For all of the subsequent discussions, the development branch %
% \footnote{It is difficult to provide a version number as a reference for the reader to compare against since this development branch is in constant flux (and released without any version numbers to even record). The development branch has many additional features compared to the `main' branch, including several important bug fixes. That being said, as of late, most of the contributions to the development branch have been updates to or the inclusion of independent physics projects that do not affect the accuracy or speed of the topic at hand. Despite the odd fix that might be introduced here or there, we do not feel that comparing results with a version of the code different from the one we used would introduce significant differences to invalidate any of our claims.} %
of \textsc{Osiris} with the addition of our collisional ionization algorithm was used. For all the subsequent tests, the code was compiled in single precision using the Intel compiler v2022.1.2 and OpenMPI v4.1.6 on the Great Lakes HPC at the University of Michigan. We perform comparisons with the open source codes \textsc{Smilei}~\cite{derouillatSmileiCollaborativeOpensource2018} and \textsc{Epoch}~\cite{arberContemporaryParticleincellApproach2015}.

\subsection{Collisional Ionization Rates}%
\label{chap:implementation:ssec:pop_rates}%
Arguably, the most critical portion of the algorithm to get correct is the rate at which ionization occurs since every succeeding aspect of the program depends on its accuracy. Crucial to this is the requirement that we compare the output from our algorithm with an independently calculated quantity. The simulations and comparisons were designed as follows. A 1D simulation using the minimum amount of cells allowed (effectively making this what we will refer to as a 0D simulation) was initialized with a uniform profile of some neutral, ionizable species. For quadratic interpolation, this amounted to 6 grid cells. A similarly uniform electron species was also placed throughout the domain and given some initial momentum in the transverse direction. Using periodic boundary conditions is equivalent to having a continuous sheet of electrons passing over an ionizable species.

This test aims to measure how accurately we can reproduce \cref{chap:implementation:eq:bateman_eqs,chap:implementation:eq:electron_den_advance}. The rates, \textsf{R}, are a function of time in addition to the ion densities themselves, which makes solving these equations challenging without additional constraints. While the rates are a function of both the electron speed and the electron number density, we seek to keep the speed constant and let the electron density evolve as we would expect it to via \cref{chap:implementation:eq:electron_den_advance}. This is because increasing the number of electrons participating in collisional ionization can enhance the process over time, a behavior we hope to capture.

This constraint is captured in our simulations by turning off any momentum loss by the electrons due to collisional ionization and artificially injecting newly ionized electrons with the same momentum as the initial electron sheet. Although this is not physically accurate with respect to momentum conservation, our goal is to verify the ionization rate; secondary momentum effects are tested separately.

Several elements were tested in this scenario, but hydrogen would naturally be the most straightforward element to test first. Over the timescale $300/\omega_p$, with a timestep of $0.05/\omega_p$, an electron sheet of density $0.1n_0$, initialized with a momentum of 1 $m_ec$ in the \(x_{2}\) direction, flowed over ionizable, neutral hydrogen with a density of $n_0$. Newly ionized electrons were also injected with a momentum of 1 $m_ec$ in the same direction. The spatial domain of the simulation grid spanned $1\times10^{9}c/\omega_p$, which was chosen to avoid any interactions between particles in one cell from another. Field ionization was turned off and collisional ionization was the only source of ionizing the material. Additionally, the PIC pusher was turned off for all of the electrons for the duration of the simulation to ensure that the momenta of all the particles remained constant.

\begin{figure}[t]
\centering
\includegraphics[width=\textwidth]{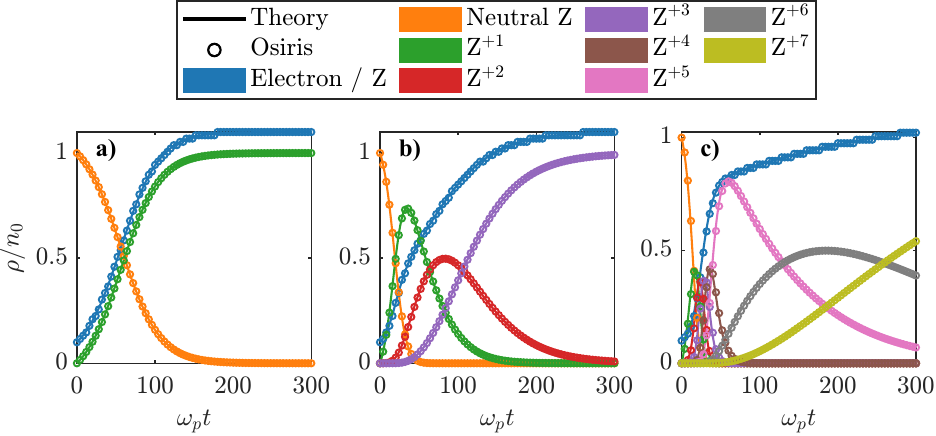}
\caption{The change in ion and electron densities for \textbf{a} hydrogen, \textbf{b} lithium, and \textbf{c} nitrogen due to collisional ionization when using the neutral object in \textsc{Osiris}.}
\label{chap:implementation:fig:neutral_rates}
\end{figure}

The results for using 64 particles per cell (ppc) are shown in \cref{chap:implementation:fig:neutral_rates}. Our theoretical values used as our `truth' are the numerical integrations of \cref{chap:implementation:eq:bateman_eqs,chap:implementation:eq:electron_den_advance} with the same initial conditions as the simulations. Encouragingly, \textsc{Osiris} can match the theoretical curves with little noise. The same simulations were run with different elements, lithium and nitrogen, with similar ideal results. The model was also validated for higher Z atoms.

\begin{figure}[t]
\centering
\includegraphics[width=\textwidth]{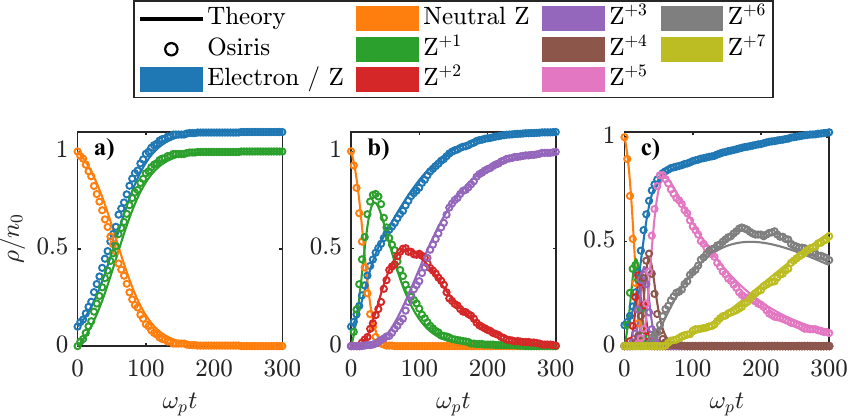}
\caption{The change in ion and electron densities for \textbf{a} hydrogen, \textbf{b} lithium, and \textbf{c} nitrogen due to collisional ionization when using the neutral object with ionizable moving ions in \textsc{Osiris}.}
\label{chap:implementation:fig:mov_ion_rates}
\end{figure}

The same test was done using the neutrals with ionizable moving ions. Similarly, the goal of this test was to verify that \cref{chap:implementation:eq:zeta_advance} is sufficient to accurately model the change in the ion number densities. The same simulation previously described was also carried out for the ionizable neutral object. Results for the same elements are shown in \cref{chap:implementation:fig:mov_ion_rates}. As with immobile ions, we accurately match our anticipated ionization rates and balance the in and outflow of ion densities with our different advancement scheme.

\begin{figure}[t]
\centering
\includegraphics[width=\textwidth]{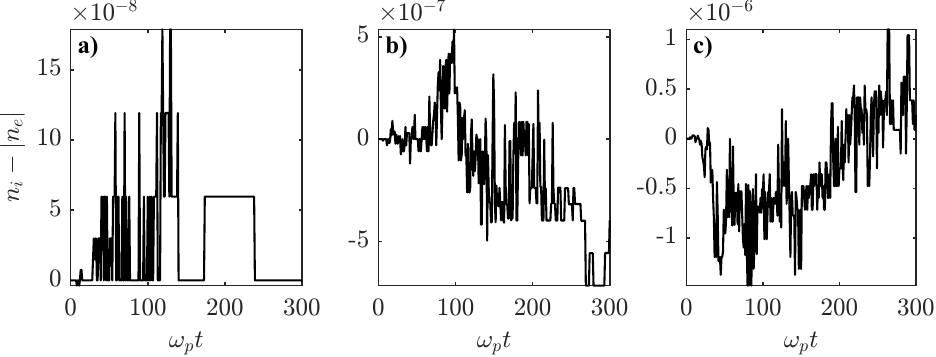}
\caption{The electron charge density subtracted from the cumulative ion charge density in our \textbf{a} hydrogen, \textbf{b} lithium, and \textbf{c} nitrogen simulations using ionizable moving ions to ensure that charge is conserved as we create, delete, and modify simulation macro-particles.}
\label{chap:implementation:fig:mov_ion_charge_cons}
\end{figure}

Unique with ionizable moving ions is the creation and destruction of macro-particles and the direct manipulation of their attributes to maintain this ionization balance. Explicitly, ion macro-particles are constantly deleted and recreated while their charge (weight) is adjusted as they are promoted from one ionization level to the next. This is in addition to creating new electron macro-particles of appropriate charge. As such, it is essential to ensure that charge is conserved throughout. \Cref{chap:implementation:fig:mov_ion_charge_cons} shows the total electron charge subtracted from the total ion charge, summed over all the ionization states, as the ionization process occurs. These diagnostics were saved in single precision, and seeing that this subtraction hovers around that level of machine precision is expected.

\subsection{Momentum Loss}%
\label{chap:implementation:ssec:ploss}%
It is also important to ensure that the momentum lost by incident macro-particles follows an anticipated physical trend. During every timestep, if a macro-particle contributes to collisional ionization, it inherently loses some of its original momentum. Physically, this can be equated to a stopping power; eventually, an electron that continuously causes collisional ionization would lose all of its momentum and come to a stop. We look to see how well our momentum loss implementation compares against a well-known stopping power formulation.

We define \(Q\) as the kinetic energy loss per unit length, or \(Q\equiv{dT}/{ds}\). This quantity is proportional to the density of the material we are interested in, \(n_{i}\), so it is often more convenient to express this as \(Q/n_{i}\), typically reported in units of MeV$\,$cm$^2$.

For cold materials (i.e., we neglect the presence of free electrons and assume that atoms are separated enough that we only collide with one at a time), \cite{rohrlichPositronElectronDifferencesEnergy1954} derived an expression for the energy lost to a not-very-dense material. Considering quantum and nuclear effects when the incident particle energy is less than that of the nuclear potential and the opposite case where the incident particle is very energetic, we can treat these collisions as regular \textit{e-e} collisions. They derived the following expression
\begin{equation}
  \frac{Q}{n_{i}}=Q_{0}\frac{Z}{\beta^{2}}\left[\ln\left(\frac{\gamma+1}{2}\frac{T^{2}}{I_{0}^{2}}\right)+\frac{1-(2\gamma-1)\ln2+(\gamma-1)^{2}/8}{\gamma^{2}}\right]\;.%
  \label{chap:implementation:eq:rohrlich}%
\end{equation}
\(Z\) is the atomic number of the atoms causing the slowing, and \(\beta\) and \(\gamma\) are the usual quantities associated with the Lorentz factor. Additionally, we define \(Q_{0}=2\pi r_{e}^{2}m_{e}c^{2}\simeq{2.55\times10^{-25}}\)~MeV$\,$cm$^2$~\cite{rohrlichPositronElectronDifferencesEnergy1954,perezEtudeTransportElectrons2011} and approximate \(I_{0}\) using the the empirical formula provided by \cite{sternheimerDensityEffectIonization1966}, \(I_{0}\simeq 9.76Z+58.8Z^{-0.19}\)~eV.

In general, the stopping power remains constant regardless of the temperature of the material. However, the relative contribution to this total stopping power from underlying physical effects does change with temperature. This is explored succinctly in \cite{perezEtudeTransportElectrons2011}. For cold materials (\(\sim\)eV), interactions between the incident electrons and electrons bound to the stopping material dominate the total contribution to the material's overall stopping power. But interactions with bound electrons are not the only physical effect to take into consideration: interactions with (already ionized) free electrons and the excitation of plasmons become particularly significant in hotter materials (\(T\gtrapprox\)100 eV). In short, the degree to which the stopping material is ionized affects its average ionization potential, \(I_{0}\), requiring a re-derivation of our stopping power expression, which is done in~\cite{solodovStoppingPowerRange2008}. Additionally, the amount of energy electrons lose to plasmons and free electrons greatly increases as the temperature of the material increases~\cite{pinesCollectiveDescriptionElectron1952} while losses to bound electrons decrease. This overall shift in contributions to the stopping power keeps it relatively constant over a wide range of temperatures. Because the stopping power is unaffected by the temperature of the material, we only focus on the case of a cold solid. %

We compare the accuracy of our approach with the anticipated result from Ref.~\cite{rohrlichPositronElectronDifferencesEnergy1954} using hydrogen and aluminum. We also looked at the accuracy of our approach using different initial energies of the electrons since \cref{chap:implementation:eq:rohrlich} is dependent on the incident particle's kinetic energy. The simulations were designed as follows: a 1D simulation was initialized with 32 grid cells and periodic boundaries. A uniform electron beam, consisting of 32 particles per cell, spanned the entire simulation domain over a similarly uniform, ionizable background species. This test was run twice using hydrogen and aluminum, with the relative densities of these species being 2:1 hydrogen/aluminum to electrons. In essence, we had these electrons flow through a persistent background of atoms as they continued to ionize and lose energy. Field ionization was turned off for these simulations, and we turned off the pusher for the initial and ionized electron particles. Hence, their only change in momentum came from losses due to collisional ionization.

\begin{figure}[t]
\centering
\includegraphics[width=\textwidth]{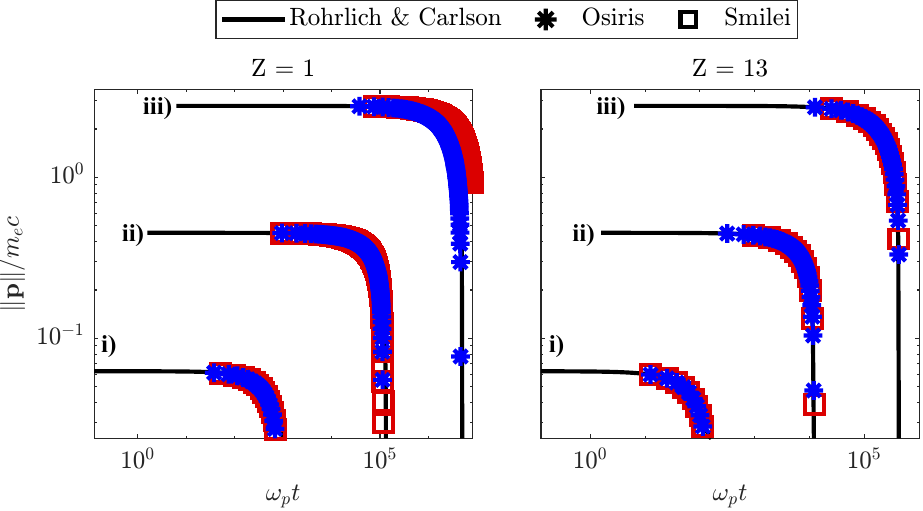}
\caption{The momentum loss of an incident electron beam due to the stopping power of both hydrogen and aluminum for three different initial electron energies of \textbf{i} 1 keV, \textbf{ii} 50 keV, and \textbf{iii} 1 MeV.}
\label{chap:implementation:fig:stopping_power}
\end{figure}

For a theoretical comparison, an electron was initialized with a kinetic energy, \(T\). Then, we iteratively decreased the electron's energy from one timestep, \(m\), to the next, \(m+1\), using \cref{chap:implementation:eq:rohrlich} via \(T^{m+1}=T^{m}-(n_{i}|v_{e}^{m}|\Delta t\frac{Q}{n_{i}})\), where \(n_{i}\) is the number density of the media the electron is traveling through and \(|v_{e}^{m}|\) is the speed of the electron at timestep \(m\). The results are shown in \cref{chap:implementation:fig:stopping_power}. We ran three simulations giving the electrons an initial energy of 1 keV, 50 keV, and 1 MeV. Our energy removal routines due to collisional ionization could accurately match the theoretical stopping distances for each initial condition. We also ran these same tests using the native implementation within \textsc{Smilei} and obtain good agreement between the codes.

\subsection{Momentum Transferred}%
\label{chap:implementation:ssec:ptransf}%
Finally, we test whether or not we prescribe the correct average momentum to newly ionized electron macro-particles. Because the algorithms for calculating the transferred momentum are slightly different depending on whether we use immobile or mobile ions, we break our discussion into either case. However, the designs of these tests were identical for both. A simulation domain of 40 cells was populated with a uniform neutral element. Over this, we had a monoenergetic electron species, with a density ratio of 1 electron per 2 ions, flow across the neutral, causing the ionization. Only the initial electron beam was allowed to ionize the background material, and neither field ionization nor effects from the secondary electrons were considered. Additionally, the particle pusher was turned off for each of these simulations. A scan of one hundred simulations was performed, varying the kinetic energy of the incident electron beam in each case. The simulation duration in all tests was sufficient to cause 100\% ionization to the maximum ionization state by the end. We ran this test for two elements, hydrogen and carbon, and various particles per cell and timesteps.

\subsubsection{Immobile Ions}%
\label{chap:implementation:sssec:ptransf_test_immobile}%
For a simple atom like hydrogen or any hydrogenic element, one particle per cell is sufficient to model the expected theoretical behavior. We encounter noticeable errors, however, when we meet the condition \(\mathsf{R}\Delta t\lessapprox1\). In this domain, our second-order finite-differencing scheme overestimates the amount of ionization that occurred. As a result, we deposit more momentum than we should to the momentum transfer grids. The coefficient we include, \(\mathsf{C}_{\text{inj}}\), to correct for small timesteps, is partially responsible for this overestimation, so increasing the number of particles per cell helps converge to the anticipated result.

\begin{figure}[t]
\centering
\includegraphics[width=\textwidth]{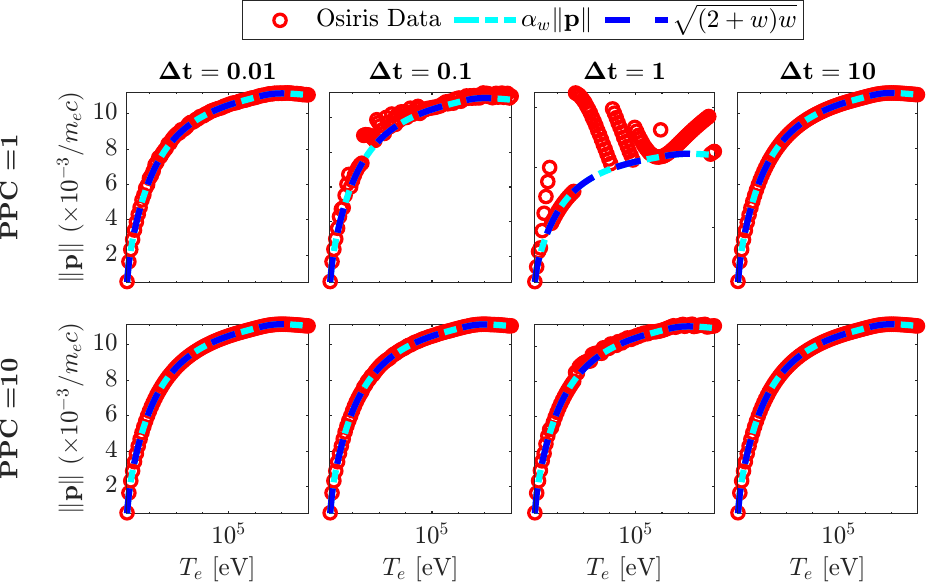}
\caption{We calculate the theoretical average momentum transferred to newly created macro-particles as a function of incident electron energy using our algorithm when ionizing hydrogen. \(\alpha_{w}|\vec{p}|\) and \(\sqrt{(2+w)w}\) are two different means of calculating how much momentum should be transferred to a newly created electron through collisional ionization. We compare these curves to the output from \textsc{Osiris}, where the momentum plotted is averaged over all the ionized electrons. \textsc{Osiris} is fairly accurate with large numbers of particles and in simulations with very small and very large timesteps. Only in the region where \(\mathsf{R}\Delta t\lessapprox1\) do we see the apparent errors in the momentum of newly ionized electrons.}
\label{chap:implementation:fig:h_ptransf_immobile}
\end{figure}
\begin{figure}[!htbp]
\centering
\includegraphics[width=\textwidth]{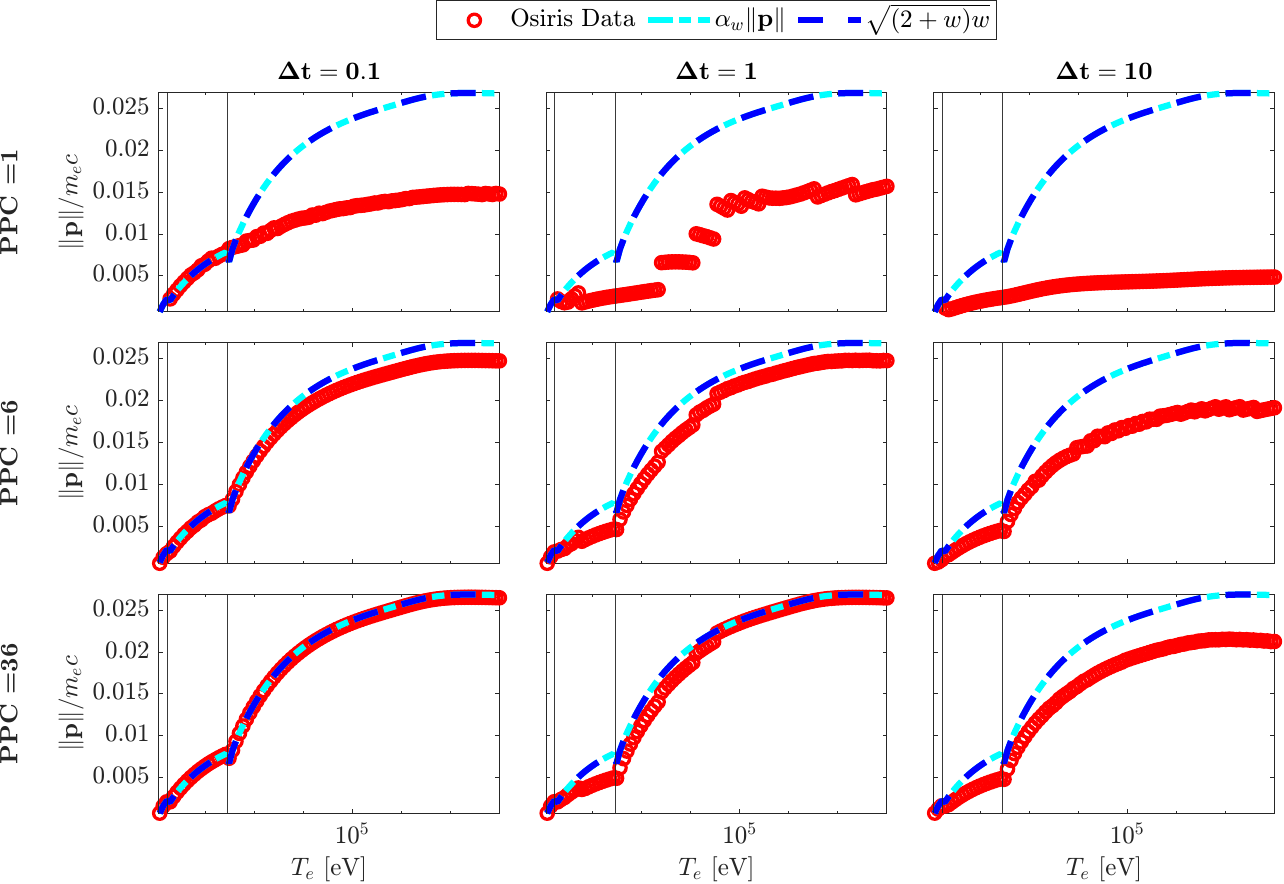}
\caption{We calculate the theoretical average momentum transferred to newly created macro-particles as a function of incident electron energy using our algorithm when ionizing carbon. \(\alpha_{w}|\vec{p}|\) and \(\sqrt{(2+w)w}\) are two different means of calculating how much momentum should be transferred to a newly created electron through collisional ionization. We compare these curves to the output from \textsc{Osiris}, where the momentum plotted is averaged over all the ionized electrons. Only when we start using multiple electron macro-particles per charge state do we start converging to the expected result. The vertical lines mark the binding energies for different atomic shells.}
\label{chap:implementation:fig:c_ptransf_immobile}
\end{figure}

%{\color{red}
Another concern is when \(z_{\text{max}}>1\) and there are low numbers of particles-per-cell, since the number of particles-per-cell is defined for all ionization states, e.g. for carbon if $z_{\text{max}}=6$ and the number of particles-per-cell is 2, as a pathological example, from \cref{chap:implementation:eq:ninj}, the only electron macro-particle gets created once the neutral has been ionized to a minimum charge state of +3. In this case, all of the momenta stored thus far in the simulation gets transferred to that one particle, summed over states. 
% }
This consequence manifests itself in \cref{chap:implementation:fig:c_ptransf_immobile}. In the first row of this figure, where we only used one particle per cell, the upper bound to the momentum assigned to ionized electrons is what we would expect the result to be if we limited the ionization of the carbon species to +3 instead of +6. This is the case for the simulations using a timestep of 0.1 and 1 (although the simulation using a timestep of 1 also has some numerical instabilities from our finite-differencing scheme). For a larger time step of 10, we do not accumulate enough momentum before we reach our injection criteria, and we underestimate the energy of ionized electrons. The simulation results converge towards what we expect only by increasing the particles per cell to at least one particle per ion charge. Note that these cases are not really an issue in practice. The user must just ensure that the ionization rate is well resolved temporally and that there are sufficient particles (per cell) to represent the ionization states involved. 

\subsubsection{Mobile Ions}%
\begin{figure}[t]
\centering
\includegraphics[width=\textwidth]{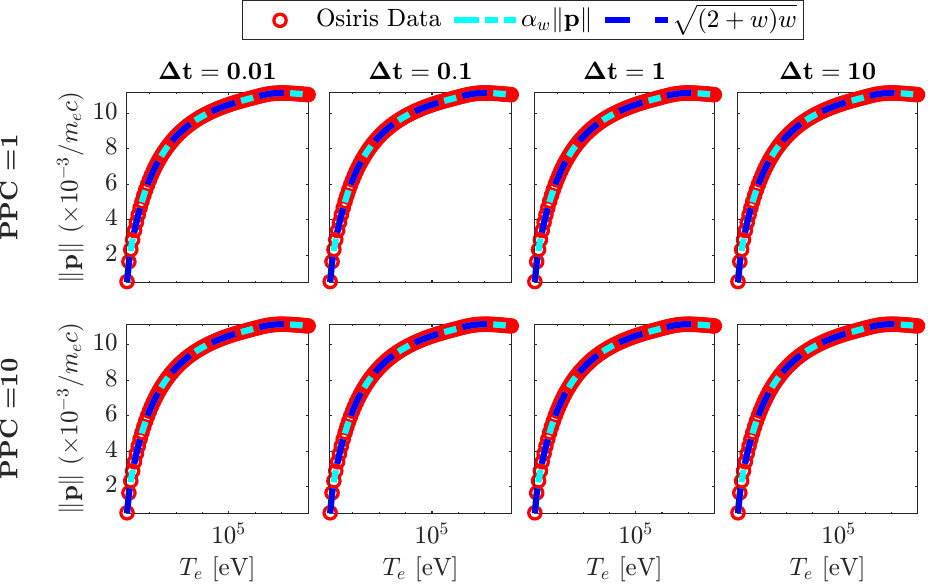}
\caption{We calculate the theoretical average momentum transferred to newly created macro-particles as a function of incident electron energy using our algorithm when ionizing hydrogen. \(\alpha_{w}|\vec{p}|\) and \(\sqrt{(2+w)w}\) are two different means of calculating how much momentum should be transferred to a newly created electron through collisional ionization. We compare these curves to the output from \textsc{Osiris}, where the momentum plotted is averaged over all the ionized electrons. We see that regardless of timestep or particles per cell, \textsc{Osiris} can exactly produce the expected momenta.}
\label{chap:implementation:fig:h_ptransf_mobile}
\end{figure}

\begin{figure}[!htbp]
\centering
\includegraphics[width=\textwidth]{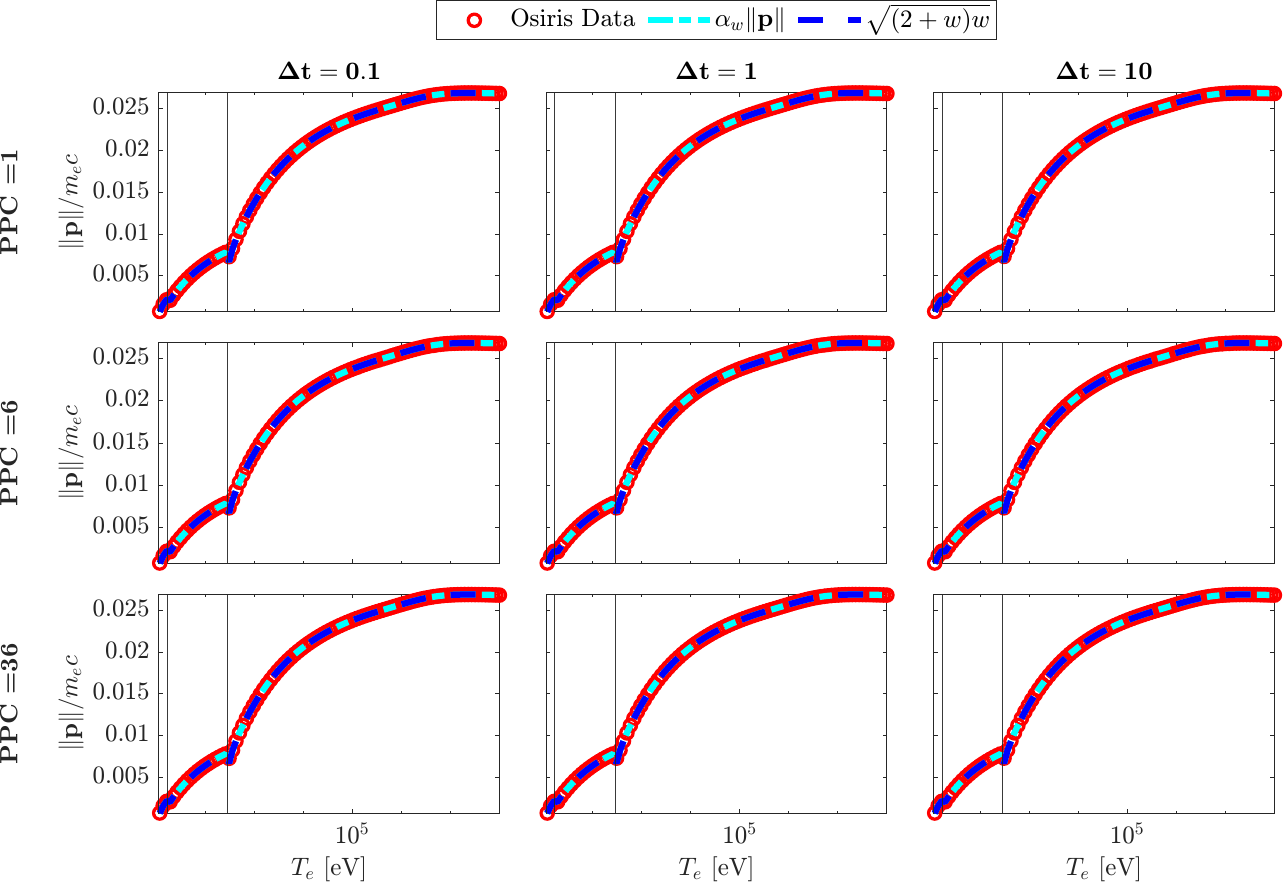}
\caption{We calculate the theoretical average momentum transferred to newly created macro-particles as a function of incident electron energy using our algorithm when ionizing carbon. \(\alpha_{w}|\vec{p}|\) and \(\sqrt{(2+w)w}\) are two different means of calculating how much momentum should be transferred to a newly created electron through collisional ionization. We compare these curves to the output from \textsc{Osiris}, where the momentum plotted is averaged over all the ionized electrons. We see that regardless of timestep or particles per cell, \textsc{Osiris} can exactly produce the expected momenta. The vertical lines mark the binding energies for different atomic shells.}
\label{chap:implementation:fig:c_ptransf_mobile}
\end{figure}

\label{chap:implementation:sssec:ptransf_test_mobile}%
A critical difference between immobile and mobile ions is the fact that one, and only one, electron is generated per ion macro-particle per promotion. This guarantee allows us to calculate the exact amount of momentum that should be passed on to each newly ionized electron. Explicitly, this approach ensures that no electron macro-particle represents the ionization of multiple charge states. We know exactly which charge state the ion started at and its charge state upon promotion, allowing us to accurately calculate the correct momenta to give to its ionized electron. \Cref{chap:implementation:fig:h_ptransf_mobile,chap:implementation:fig:c_ptransf_mobile} show the momentum transferred to ionized electrons for two different Z material, hydrogen and carbon, for the same set of timesteps we used in the immobile case. In every test, there is good agreement with theory.

\section{Code Validation}%
\label{chap:implementation:sec:comparisons}%
At the time of writing this document, for benchmarking we used the most current release of \textsc{Epoch}, v4.19, which has their most updated collisional ionization implementation~\cite{morrisImprovementsCollisionalIonization2022}. We compiled \textsc{Epoch} using the Intel compiler v2022.1.2 and OpenMPI v4.1.6 on the Great Lakes HPC at the University of Michigan. Beyond the default compiler options, we used the following compiler flags: -O3 -heap-arrays 64 -ipo -xHost, and activated the following pre-processor defines provided with the code: \verb|NO_TRACER_PARTICLES|, \verb|NO_PARTICLE_PROBES|, \verb|PREFETCH|. We also used \textsc{Smilei} v4.7 compiled using the Intel compiler v2022.1.2, Intel MPI compiler v2021.5.1, and Python v3.10.9. Additionally, we used the machine-specific compiler flags \textsc{Smilei} provides for Skylake processors (the processors used in Great Lakes) and Intel compilers.
\subsection{Benchmarking with \textsc{Smilei} and \textsc{Epoch}}
Considering the differences in the approaches between our finite difference deterministic algorithm and the Monte-Carlo method of \textsc{Smilei} and \textsc{Epoch}, in addition to benchmarking and validating our code, we were also interested in the relative accuracy and scaling. The first comparison to make is simply integrating the ionization rate with the different algorithms. 

\begin{figure}[t]
\centering
\includegraphics[width=\textwidth]{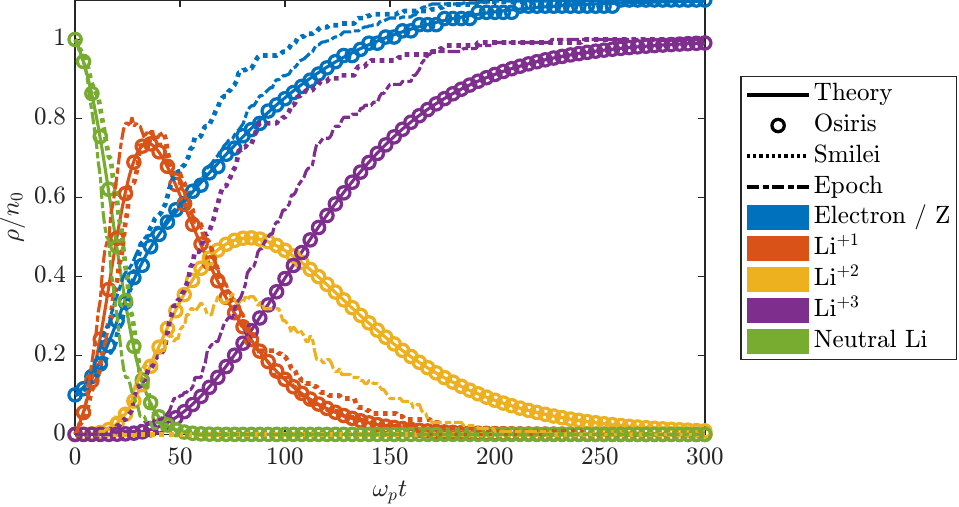}
\caption{A comparison of the electron and ionization state number densities of lithium using the three PIC codes \textsc{Osiris}, \textsc{Smilei}, and \textsc{Epoch}. The procedure outlining this test was consistent between all three and described in \cref{chap:implementation:ssec:pop_rates}.}
\label{chap:implementation:fig:all_pic_li_rate_compare}
\end{figure}

Both \textsc{Smilei} and \textsc{Epoch} do not offer control over certain aspects of their collisional ionization routines as a standard feature. To directly compare these codes to \textsc{Osiris}, however, we modified each of them to both stop the removal of energy from ionizing electrons and inject new electrons with specific momenta in line with our procedure outlined in \cref{chap:implementation:ssec:pop_rates} to isolate the rate calculation. 
The results of these tests using a lithium neutral species are shown in \cref{chap:implementation:fig:all_pic_li_rate_compare}. The results from all three codes, \textsc{Smilei}, \textsc{Epoch} and \textsc{Osiris} are in reasonable agreement. Some of the slight discrepancies between the codes are simply attributed to differences in the physical cross sections used by each code. \textsc{Smilei} uses the MRBEB model presented in \cite{kimExtensionBinaryencounterdipoleModel2000}, and \textsc{Epoch} uses the MBELL model~\cite{bellRecommendedDataElectron1983} for \(Z\leq18\) and the MRBEB model for the rest. Both codes use slightly different values for the binding energies. However, considering these different cross sections does not fully account for the deviations we see. For example, \textsc{Smilei} implemented a cumulative probability in their Monte-Carlo scheme that considers a macro-particle undergoing multiple ionization events within a single timestep \cite{derouillatSmileiCollaborativeOpensource2018}, a scheme originally derived by \cite{nuterFieldIonizationModel2011}. The initial implementation of this multiple ionization was presented \emph{only} in the context of field ionization and developed to address the following scenario. Often, PIC simulations using field ionization routines were limited in the timesteps one could use; if the local electric field was too strong and the timestep too large, the previous implementation would allow only a single electron to be ionized rather than the multiple we would expect. This constrains users to use a smaller timestep to capture the ionization dynamics properly. Thus, this cumulative probability was intended to relieve this numerical constraint.

\begin{figure}[t]
\centering
\includegraphics[width=\textwidth]{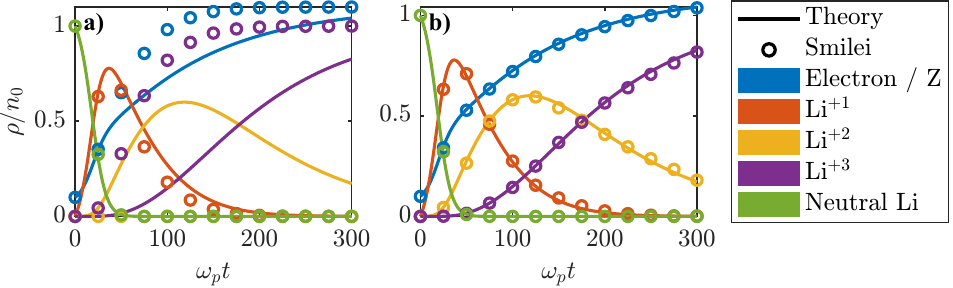}
\caption{Ionization state and electron number densities of lithium using \textsc{Smilei} \textbf{a} with multiple ionizations per timestep and \textbf{b} without multiple ionizations per timestep. The theory lines presented here are the numerically integrated solutions to \cref{chap:implementation:eq:bateman_eqs,chap:implementation:eq:electron_den_advance} utilizing the cross sections used within \textsc{Smilei}.}
\label{chap:implementation:fig:smilei_mult_ion_compare}
\end{figure}

This formulation works well for field ionization but is erroneous when extended to collisional ionization. The cross-section calculations we have presented, a form of which is used within \textsc{Smilei}, are specific for single interactions between one incident electron and an ion, i.e., a single ionization event. Double or multiple ionization cross sections exist but are orders of magnitude smaller than their single ionization counterparts~\cite{stephanMassSpectrometricDetermination1980}.
%One could argue that, within the PIC infrastructure, these macro-particles represent a collective of many electrons and ions. Thus, the concept of multiple ionizations occurring within a single timestep is reasonable if you assume that an ion interacts with several particles of the collective. However, the problem still 
The problem arising from the multiple ionization calculation of collisional ionization rates is that inherently, after a single ionization event, the number densities of the electrons and ions change, thus changing the ionization rates during the process. Within this multiple ionizations per timestep scheme, the relative species' densities, and consequently the rates used to calculate the probability of ionization, are \emph{not} updated within the same timestep, which is a source of inaccuracy. 
%Considering that the incident and ionized electron leave the ion within $10^{-16}$~s of the interaction~\cite[chap.~5.1.1]{markElectronImpactIonization1985}, accounting for the changes in densities seems particularly important when considering multiple ionizations within a single timestep.
A comparison between using and not using this approach is shown in \cref{chap:implementation:fig:smilei_mult_ion_compare}. Beyond inaccuracies in the solutions compared to \cref{chap:implementation:eq:bateman_eqs,chap:implementation:eq:electron_den_advance}, this method is explicitly unable to capture the presence and evolution of any Li\(^{+2}\) when allowing for a macro-particle to ionize multiple times per timestep. Turning off this multiple ionization procedure (with a small edit to the code) results in the expected behavior and agreement between the codes.

\begin{figure}[t]
\centering
\includegraphics[width=\textwidth]{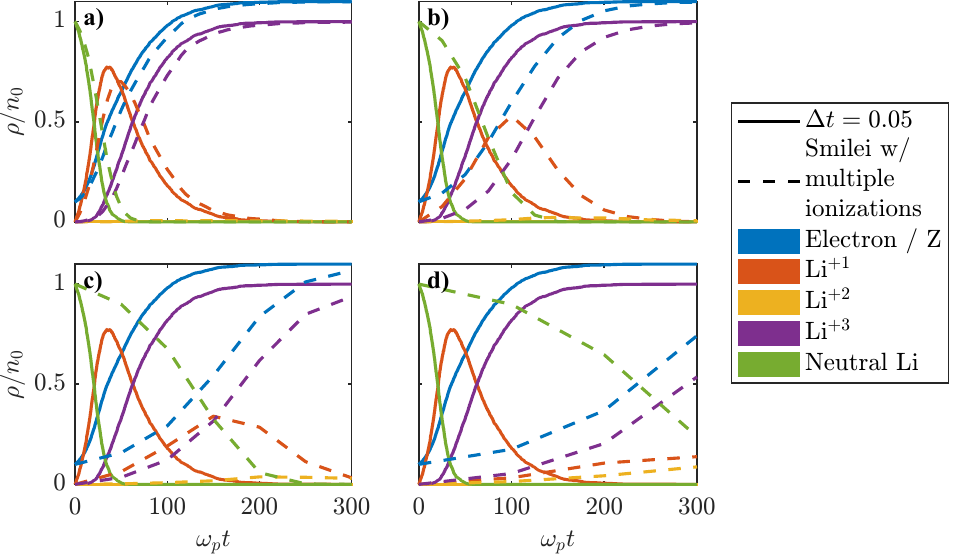}
\caption{Results from \textsc{Smilei} showing the change in the number density of the ionization states of lithium and electrons present from simulations using a timestep of \textbf{a}~\(\Delta t=5/\omega_p\), \textbf{b}~\(\Delta t=25/\omega_p\), \textbf{c}~\(\Delta t=50/\omega_p\), and \textbf{d}~\(\Delta t=100/\omega_p\). These results are from the unedited collisional ionization algorithm in \textsc{Smilei} that allows a macro-particle to be ionized multiple times within a single timestep. The comparison curves presented are also the results from \textsc{Smilei} but using a timestep of $0.05/\omega_p$, where the results have reached their converged value.}
\label{chap:implementation:fig:smilei_w_mult_ion_dt_scan}
\end{figure}

\begin{figure}[t]
\centering
\includegraphics[width=\textwidth]{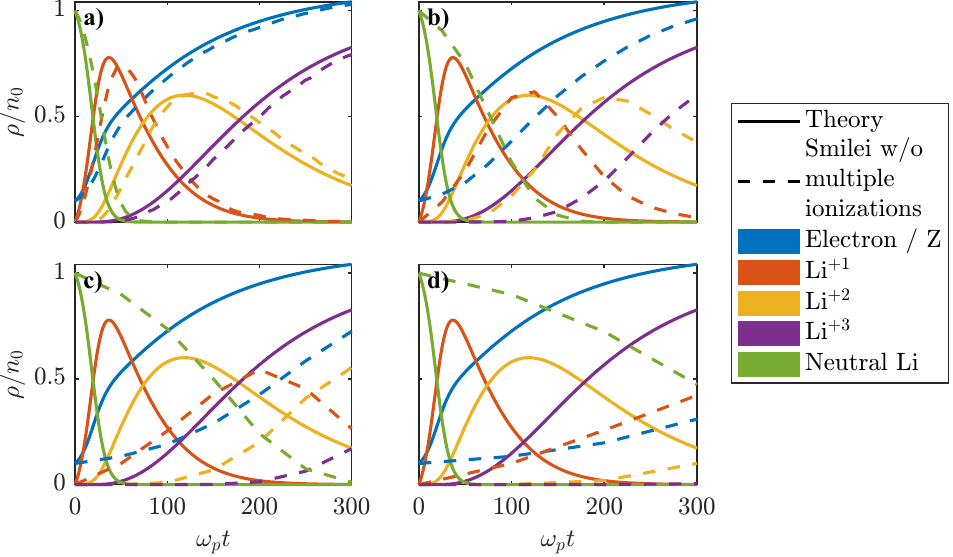}
\caption{Results from \textsc{Smilei} showing the change in the number density of the ionization states of lithium and electrons present from simulations using a timestep of \textbf{a}~\(\Delta t=5/\omega_p\), \textbf{b}~\(\Delta t=25/\omega_p\), \textbf{c}~\(\Delta t=50/\omega_p\), and \textbf{d}~\(\Delta t=100/\omega_p\). These results are from an edited version of the collisional ionization algorithm in \textsc{Smilei} that does not allow a macro-particle to be ionized multiple times in a single timestep. The theory curves we compare our results against are the numerically integrated solutions to \cref{chap:implementation:eq:bateman_eqs,chap:implementation:eq:electron_den_advance} using the cross sections present in \textsc{Smilei}.}
\label{chap:implementation:fig:smilei_wo_mult_ion_dt_scan}
\end{figure}

%In theory, the advantage of this approach would be allowing larger timesteps in our simulations, thus saving computation time. 
\Cref{chap:implementation:fig:smilei_w_mult_ion_dt_scan}, shows the ionization for different timescales. %Since \textsc{Smilei} does not provide an expression for the theoretical evolution of their species using this multiple ionization scheme, W
Here, we compare how accurate the ionization rates from \textsc{Smilei} are when using a small timestep (in this case, a well-converged result using a timestep of $0.05/\omega_p$) versus various larger time steps. The number densities of these different species are underestimated. If we turn off these multiple ionizations, such that the implementation is more in line with the particle-particle algorithm, the agreement is improved, \cref{chap:implementation:fig:smilei_wo_mult_ion_dt_scan}. For this figure, the results from \textsc{Smilei} are compared with solutions to the differential rate equations, \cref{chap:implementation:eq:bateman_eqs,chap:implementation:eq:electron_den_advance} but using \textsc{Smilei}'s cross sections.

\clearpage

\begin{figure}[t]
\centering
\includegraphics[width=\textwidth]{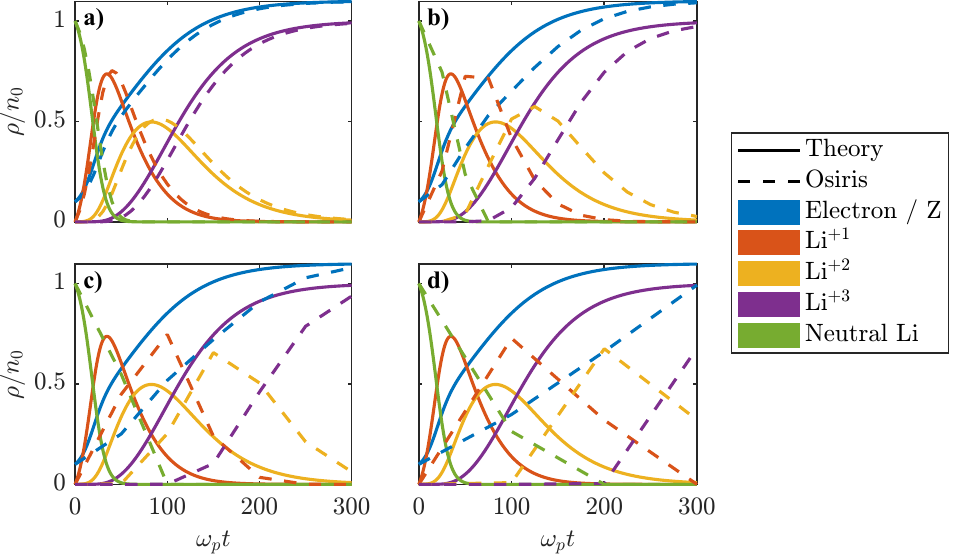}
\caption{Results from \textsc{Osiris} showing the change in the number density of the ionization states of lithium and electrons present from simulations using a timestep of \textbf{a}~\(\Delta t=5/\omega_p\), \textbf{b}~\(\Delta t=25/\omega_p\), \textbf{c}~\(\Delta t=50/\omega_p\), and \textbf{d}~\(\Delta t=100/\omega_p\). The theory curves we compare our results against are the numerically integrated solutions to \cref{chap:implementation:eq:bateman_eqs,chap:implementation:eq:electron_den_advance} using our cross sections.}
\label{chap:implementation:fig:osiris_dt_scan}
\end{figure}

As seen in \cref{chap:implementation:fig:osiris_dt_scan}, the new algorithm in \textsc{Osiris} compares well with the results of \textsc{Smilei} with the multiple ionization event effect switched off. It remains relatively accurate when using a timestep of $5/\omega_p$, which is large in relation to the timestep of $0.05/\omega_p$ used in all the previous results presented. The results start to deviate for the larger timesteps of $25/\omega_p$ and $50/\omega_p$, but the deviation is smaller compared to \textsc{Smilei}. 
Stopping power tests were also run in \textsc{Smilei}. These results were previously shown in \cref{chap:implementation:fig:stopping_power} where \textsc{Smilei} agreed with both \textsc{Osiris} and the theory curve.

% \begin{figure}[t]
% \centering
% \includegraphics[width=\textwidth]{figs/mov_ion_ppc_compare.pdf}
% \caption{The number of electron and ion macro-particles created over time when using ionizable moving ions in \textsc{Osiris}, \textsc{Smilei}, and \textsc{Epoch}. These were recorded specifying for each species to have \textbf{a} 10, \textbf{b} 100, \textbf{c} 1000, and \textbf{d} 10000 ppc. The number of electron macro-particles in \textsc{Osiris} is double that compared to the other PIC codes because both the initial electrons that start the ionization process and the newly injected ionized electrons (two separate species within \textsc{Osiris}) were each specified to have the same ppc. Should we have halved the number of particles we specified for each species, the lines for \textsc{Osiris} would align with the other PIC codes.}
% \label{chap:implementation:fig:mov_ion_ppc_compare}
% \end{figure}
 
\subsection{Execution time}
\begin{figure}[t]
\centering
\includegraphics[width=\textwidth]{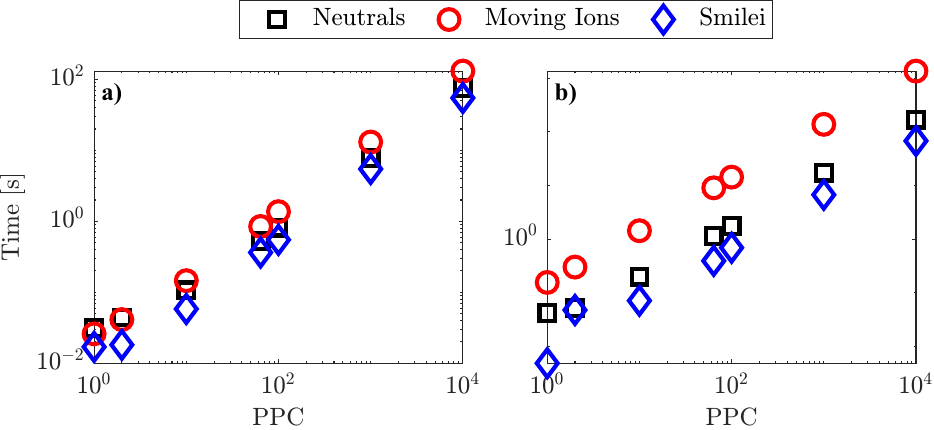}
\caption{Code execution time for the collisional ionization models in \textsc{Osiris} and \textsc{Smilei} versus the number of particles per cell used. For \textsc{Osiris}, we display the execution time for our collisional ionization module using both the base neutral object and the neutral with ionizable moving ions. We show the differences in execution time for a low and high (comparatively) Z element, \textbf{a} lithium and \textbf{b} argon. These execution numbers were taken from the ionization rate comparison studies.}
\label{chap:implementation:fig:timing_comparison}
\end{figure}

For benchmarking the execution time of our \textsc{Osiris} algorithm, we compared with the runtime of \textsc{Smilei}. These two codes provided the ability to time specific portions of a PIC loop, allowing us to see the timing of the collisional ionization routines specifically; \textsc{Epoch} only provides the cumulative run time for an entire simulation. The previously discussed rate calculations were timed, and the results are plotted in \cref{chap:implementation:fig:timing_comparison}. The collisional ionization routines in both \textsc{Smilei} and \textsc{Osiris} scale at the same rate, with the scaling with particles per cell being linear, as expected. However, our implementation in \textsc{Osiris} is slower in all cases. While we performed basic optimization of our implementation in \textsc{Osiris}, in the form of data structures that helped localize data in memory for faster access times and the organization of certain calculations such that data used in successive loop iterations remained in CPU cache as long as possible, more optimization could improve the speed.

Some improvements can be made to rectify this. Within \textsc{Osiris} are vectorized (SIMD or single instruction multiple data) implementations of the current deposition and pusher routines: routines written using specific machine instructions for processors that provide them. Given the similarities between our implementation and these other portions of the code, it is not hard to imagine an extension of these vectorized algorithms to work with our collisional ionization module, hopefully decreasing its run time.

The actual implementation of our algorithm was also developed to minimize refactoring of the existing architecture within the code base. So, a perhaps less sophisticated solution than that previously mentioned would be a more careful integration of our algorithm into the core PIC loop.

\subsection{Convergence and accuracy}%
\label{chap:implementation:sec:comments}%
%

% However, a closer look at \cref{chap:implementation:fig:all_pic_li_rate_compare} can open a more interesting conversation. Ignoring the already discussed deviations in each implementation from the theoretical model, we have yet to discuss the amounts of noise generated with each approach.

\begin{figure}[!htb]
\centering
\includegraphics[width=0.8\textwidth]{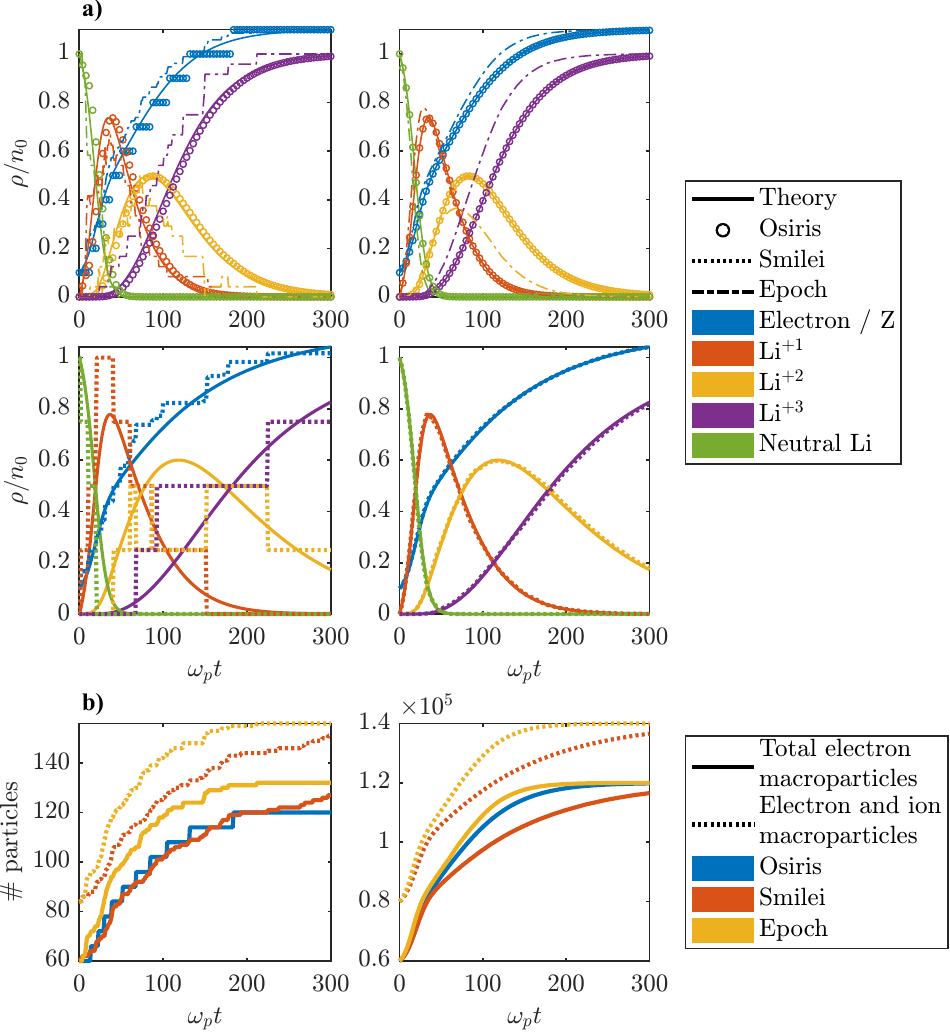}
\caption{\textbf{a} The change in ion densities due to only collisional ionization in lithium using the codes \textsc{Osiris} and \textsc{Epoch} (top) and \textsc{Smilei} (bottom). These simulations were run using ten ppc (left) and 10,000 ppc (right) to show the accuracy of each implementation using a low particle count and their convergence to their respective solutions when using larger particle counts. In these simulations, the theory curves are calculated using the cross sections from \textsc{Osiris} for the top simulations and the cross sections from \textsc{Smilei} without multiple ionizations per timestep for the bottom simulations. \textbf{b} The growth in the number of macro-particles in each code when using ten ppc (left) and 10,000 ppc (right) (each simulation uses six cells). Ions were not allowed to move in these \textsc{Osiris} runs, so there are no ion macro-particles to count in these simulations.}
\label{chap:implementation:fig:rate_ppc_compare}
\end{figure}

In our previous results, we kept the number of macro-particles used constant. The convergence of the algorithm with particle number $N_{PPC}$ is an interesting comparison to make, since the Monte-Carlo methods would be expected to display a slow $1/\sqrt{N_{PPC}}$ convergence in statistical errors. The right two panels of \cref{chap:implementation:fig:rate_ppc_compare}a show a well converged result, where using ten-thousand ppc allows the number density of each ionization state to match their theoretical results with minimal deviation. The two panels on the left use ten particles per cell. The difference in noise between the deterministic approach and the other particle-based algorithms is apparent. \textsc{Osiris} can match the theoretical curves with great accuracy even when using a smaller number of macro-particles, although the electron density increases step-wise owing to the deterministic creation of macro-electrons. The Monte Carlo codes show expected statistical errors. Where this may be an issue is for ionization fronts by fast electrons, which in practice may have low macroparticle numbers in an expanding hot electron flow. Noise in the ionization may seed ionization filamentation instabilities. 
% \footnote{As described in \cref{chap:implementation:ssec:pop_rates}, the spatial domain of these simulations was quite large, all with the goal of preventing fields from one cell affecting its neighbors and to prevent particles from moving from cell to cell. Since there was no variation in our initial conditions, each cell follows the same ionization trend with our deterministic algorithm, so taking an average of the results across cells does not provide us any benefit. To have an accurate comparison, though, this would mean that we only want to look at the results in the rates of ionization in a single cell from our \textsc{Smilei} simulations. Thus, the results we present here are the output of a randomly picked cell from the \textsc{Smilei} simulations.}

In \cref{chap:implementation:fig:rate_ppc_compare}b, we compare the total number of particles generated in each code for both scenarios. The takeaway is that \textsc{Osiris} can produce the level of accuracy we present while creating the same number of particles as the other codes. (In fact, we achieve this accuracy using less \emph{total} resources than \textsc{Smilei} when you account for the ion macro-particles that get created as well, but this has the caveat that ions do not move in this situation for \textsc{Osiris}. Averaging the results from \textsc{Smilei} over all grid cells renders a more accurate result but is not a one-to-one comparison with our results from \textsc{Osiris}.)

The step-like feature of the electron density from the \textsc{Osiris} results is not an inherent consequence of how we conduct ionization but more a consequence of the criteria we use to inject new macro-particles into the simulation. As a reminder, a certain amount of charge must be generated to warrant the injection of a new particle, which depends on the particles per cell used. The ion densities are continuous grid values and do not have the same limitation. We described this aspect with more detail in \cref{chap:implementation:ssec:part_inject}.

\begin{figure}[!htbp]
\centering
\includegraphics[width=\textwidth]{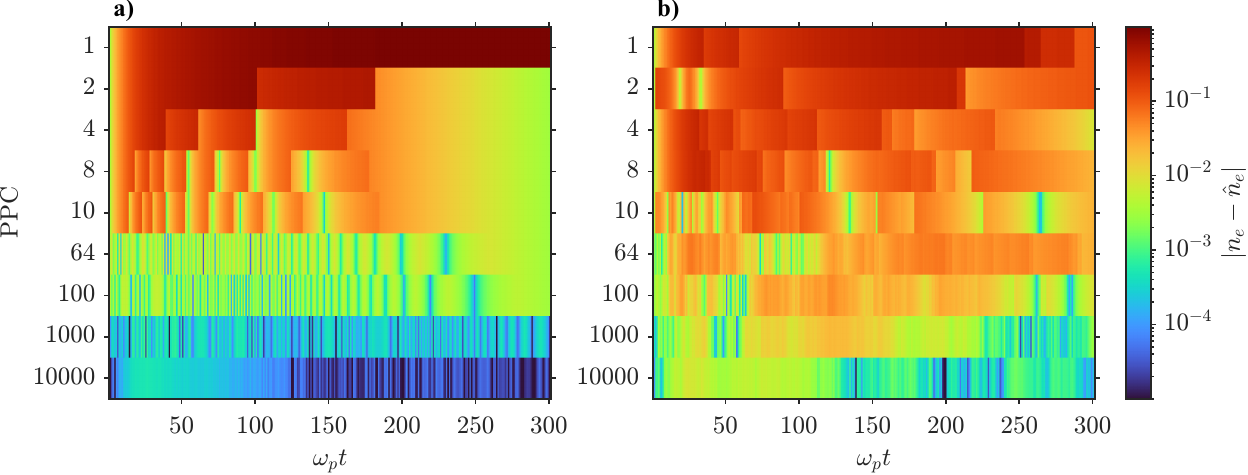}
\caption{A comparison of the residuals for the electron density in \textbf{a} \textsc{Osiris} and \textbf{b} \textsc{Smilei} for various particle counts.}
\label{chap:implementation:fig:e_ppc_residual}
\end{figure}

\begin{figure}[!htbp]
\centering
\includegraphics[width=0.8\textwidth]{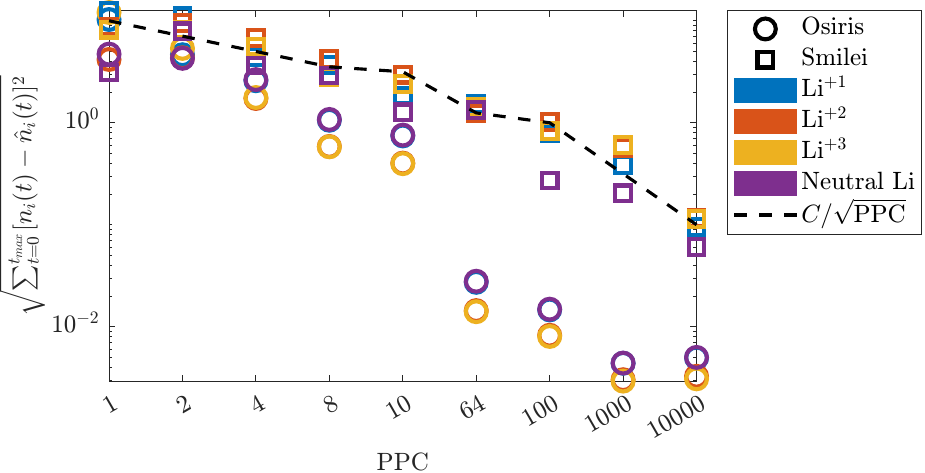}
\caption{The standard deviation for each ionization level of lithium comparing the discrepancy between the outputs of \textsc{Osiris} and \textsc{Smilei}, \(n_{i}\), with their respective theoretical solutions, \(\hat{n}_{i}\), for various particle counts. The line \(1/\sqrt{\text{PPC}}\) is included to show the general trend in error falloff and is vertically adjusted to visually line up with the data by the constant \(C\). Here, we use a value of \(C=10\).}
\label{chap:implementation:fig:ppc_rss}
\end{figure}

Quantitatively, we display the deviations in either implementation in \cref{chap:implementation:fig:e_ppc_residual,chap:implementation:fig:ppc_rss}. \Cref{chap:implementation:fig:e_ppc_residual} shows the residuals between the electron number density, \(n_{e}\), and each code's respective theoretical value, \(\hat{n}_{e}\), throughout the entire simulation for a variety of ppc values. The residuals are comparatively high in \textsc{Osiris} for the simulations with 1 or 2 particles per cell. However, since we ran these simulations ionizing lithium, specifying either 1 or 2 particles per cell does not allow us to satisfactorily resolve each ionization state's contribution of electrons that can aid in further ionization. \textsc{Smilei}, while its residuals are still relatively high, performs better than \textsc{Osiris} in this regard. Once we reach a multiplicity of electrons per ionization state, the residuals with our deterministic algorithm quickly decrease compared to \textsc{Smilei}'s implementation.

\Cref{chap:implementation:fig:ppc_rss} shows each ionization state's standard deviation, summed over the entire simulation duration. The $1/\sqrt{N_{PPC}}$ scaling is also plotted as a dashed black line, which is the statistical error we would expect to occur within a Monte-Carlo simulation. Both \textsc{Osiris} and \textsc{Smilei} are similar in their relative error for low particle numbers. However, once there are multiple particles per ionization state, our deterministic algorithm begins to see a quicker decrease in its error. At a certain point, the algorithm in \textsc{Osiris} sees a sharp drop in its cumulative error. \textsc{Smilei} more or less follows the expected noise level. In this regard, our algorithm shows some improvement in accuracy with a moderate amount of resources.

The algorithm presented here can be extended beyond electron-ion impact ionization. In fact, with minimal modification, we extended the same design to work with incident protons and water molecules \cite{streeterStableLaseraccelerationHighflux2025}. %We only needed to calculate the relevant cross sections, so this paradigm can easily be applied to many different incident particles and targets.

\section{Conclusion}%
\label{sec:conclusion}%
In conclusion, we have demonstrated our collisional and field ionization module for the particle-in-cell framework OSIRIS. It has been verified and benchmarked against two particle-in-cell codes with Monte-Carlo algorithms. It should be noted that the overall algorithm includes only field and electron impact ionization. Processes including three-body recombination, radiative recombination, dielectronic recombination, and photoionization can become significant depending on the system’s temperature and density and are not included in the current version of the model. Reaction rates for these processes can be derived from e.g. the principle of detailed balance, and could be seamlessly incorporated into the \textsc{Osiris} framework we have developed as additional source or loss terms in the evolution equations for ion charge state densities. One limitation of the model is that an ion is assumed to be in the ground state at all times. If one were interested in the full atomic kinetics, keeping track of the charge and excitation states in general is nontrivial but could be implemented.

It should be noted that in the current implementation, the ionizable moving ions routine does not take into account a Lorentz transform to the ion's frame of reference before calculating the cross sections or momentum values due to a collisional ionization event. In part there is a fundamental problem that with this algorithm the rate and momentum transfer information is averaged over. At least on average this could be addressed by transforming to the frame of the ion fluid momentum. Another simplification, which is used in all the codes, is the assumption of collinear momentum loss and that newly created particles are initialized with collinear momentum. Experimental measurements suggest that a fast ionizing electron will indeed be almost undeflected from its original trajectory~\citep{ehrhardtCollisionalIonizationHelium1972}. The slower secondary electrons, however, are ejected with an angular distribution with respect to the direction of the incident particle, with the majority being measured at angles between 30-90$^\circ$ ~\cite{ehrhardtCollisionalIonizationHelium1972,jungAngularCorrelationOutgoing1975,mansonEnergyAngularDistribution1975,kimAngularEnergyDistributions1983,inokutiTheoryElectronDegradation1987,ruddUserfriendlyModelEnergy1989}. Functions describing these angular distributions could be sampled and a statistically chosen direction assigned to newly created macro-particles to model these experimental observations better.

Last, several extensions to our algorithmic implementation within \textsc{Osiris} warrant attention. The present parallelization strategy utilizes the MPI methodology, dividing the computational grid into subdomains managed independently by processors. While effective for many geometries, this necessitates data communication across subdomain interfaces, and excessive grid subdivision (with large processor counts) can introduce communication overhead that erodes the expected performance gains. Shared-memory parallelization presents a solution for alleviating processor communication within subdomains, enabling multiple processors to act on the same domain and thereby improving scalability.
Furthermore, optimization of ionization rate deposits and interpolation routines can benefit from machine-specific vectorization instructions. Such special vector operations permit simultaneous data processing (of multiple macro-particles, for example) rather than sequential handling by the CPU. While compilers attempt automatic vectorization conservatively, reorganizing code or explicit vector instruction implementation can achieve higher performance. The current deposition scheme in \textsc{Osiris} already integrates these parallelization improvements and may serve as an exemplar for future optimizations.

\section*{Acknowledgments}
This work was supported by the Air Force Office of Scientific Research under Grant \#FA9550-19-1-0072.
\bibliographystyle{unsrt} 
\bibliography{references}

\end{document}